\documentclass[sii,authoryear]{ipart}

\RequirePackage{hyperref}
\RequirePackage{amsthm,amsmath,amsfonts,amssymb}

\usepackage{graphicx}
\usepackage{cuted}

\usepackage{algorithm}
\usepackage{algpseudocode}

\usepackage{hyperref}

\setkeys{Gin}{width=.9\textwidth,height=18cm,keepaspectratio} 

\startlocaldefs
\endlocaldefs

\pubyear{2019}
\volume{0}
\issue{0}
\firstpage{1}
\lastpage{1}

\begin{document}

\begin{frontmatter}

\title{Smooth Online Parameter Estimation for time varying VAR models with application to rat local field potential activity data}

\begin{aug}
\author{\inits{A.~Y.B.} \fnms{Anass} \snm{El Yaagoubi Bourakna}\ead[label=e1]{anass.bourakna@kaust.edu.sa} \thanksref{t2}},
\author{\inits{M.~P.} \fnms{Marco} \snm{Pinto}\ead[label=e2]{Marco.Pinto@oslomet.no}},
\author{\inits{N.~F.} \fnms{Norbert} \snm{Fortin}\ead[label=e3]{norbert.fortin@uci.edu}}
\and
\author{\inits{H.~O.} \fnms{Hernando} \snm{Ombao}\ead[label=e4]{hernando.ombao@kaust.edu.sa}\thanksref{t2}}

\address{Statistics Program, King Abdullah University of Science and Technology, \printead{e1}, \printead{e4}}

\address{Oslo Metropolitan University, \printead{e2}}

\address{University of California Irvine, \printead{e3}}

\thankstext{t2}{Corresponding author.}

\end{aug}


\begin{abstract}
    Multivariate time series data appear often as realizations of non-stationary processes where the covariance matrix or spectral matrix smoothly evolve over time. Most of the current approaches estimate the time-varying spectral properties only retrospectively - that is, after the entire data has been observed. Retrospective estimation is a major limitation in many adaptive control applications where it is important to  estimate these properties and detect changes in the system as they happen in real-time. To overcome this limitation, we develop an online estimation procedure that gives a real-time update of the time-varying parameters as new observations arrive. One approach to modeling non-stationary time series is to fit time-varying vector autoregressive models (tv-VAR). However, one major obstacle in online estimation of such models is the computational cost due to the high-dimensionality of the parameters. Existing methods such as the Kalman filter or local least squares are feasible. However, they are not always suitable because they provide noisy estimates and can become prohibitively costly as the dimension of the time series increases. In our brain signal application, it is critical to develop a robust method that can estimate, in real-time, the properties of the underlying stochastic process, in particular, the spectral brain connectivity measures. For these reasons we propose a new smooth online parameter estimation approach (SOPE) that has the ability to control for the smoothness of the estimates with a reasonable computational complexity. Consequently, the models are fit in real-time even for high dimensional time series. We demonstrate that our proposed SOPE approach is as good as the Kalman filter in terms of mean-squared error for small dimensions. However, unlike the Kalman filter, the SOPE has lower computational cost and hence scalable for higher dimensions. Finally, we apply the SOPE method to local field potential activity data from the hippocampus of a rat performing an odor sequence memory task. As demonstrated in the video, the proposed SOPE method is able to capture the dynamics of the connectivity as the rat samples the different odor stimuli.
\end{abstract}

\begin{keyword}
    \kwd{Online parameter estimation}
    \kwd{Locally stationary processes}
    \kwd{Dynamic brain connectivity}
    \kwd{Dynamic spectral connectivity}
\end{keyword}



\end{frontmatter}

\section{Introduction}
\label{sec:intro}

\begin{figure*}
\centering
    \begin{minipage}{.5\textwidth}
      \centering
      \includegraphics[width=\linewidth]{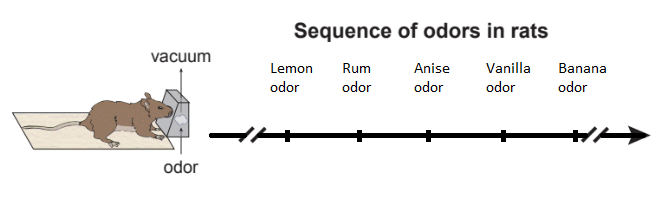}
      \caption{Odor sequence memory task.}
      \label{fig:rat_sequence}
    \end{minipage}%
    \begin{minipage}{.5\textwidth}
      \centering
      \includegraphics[width=.75\linewidth]{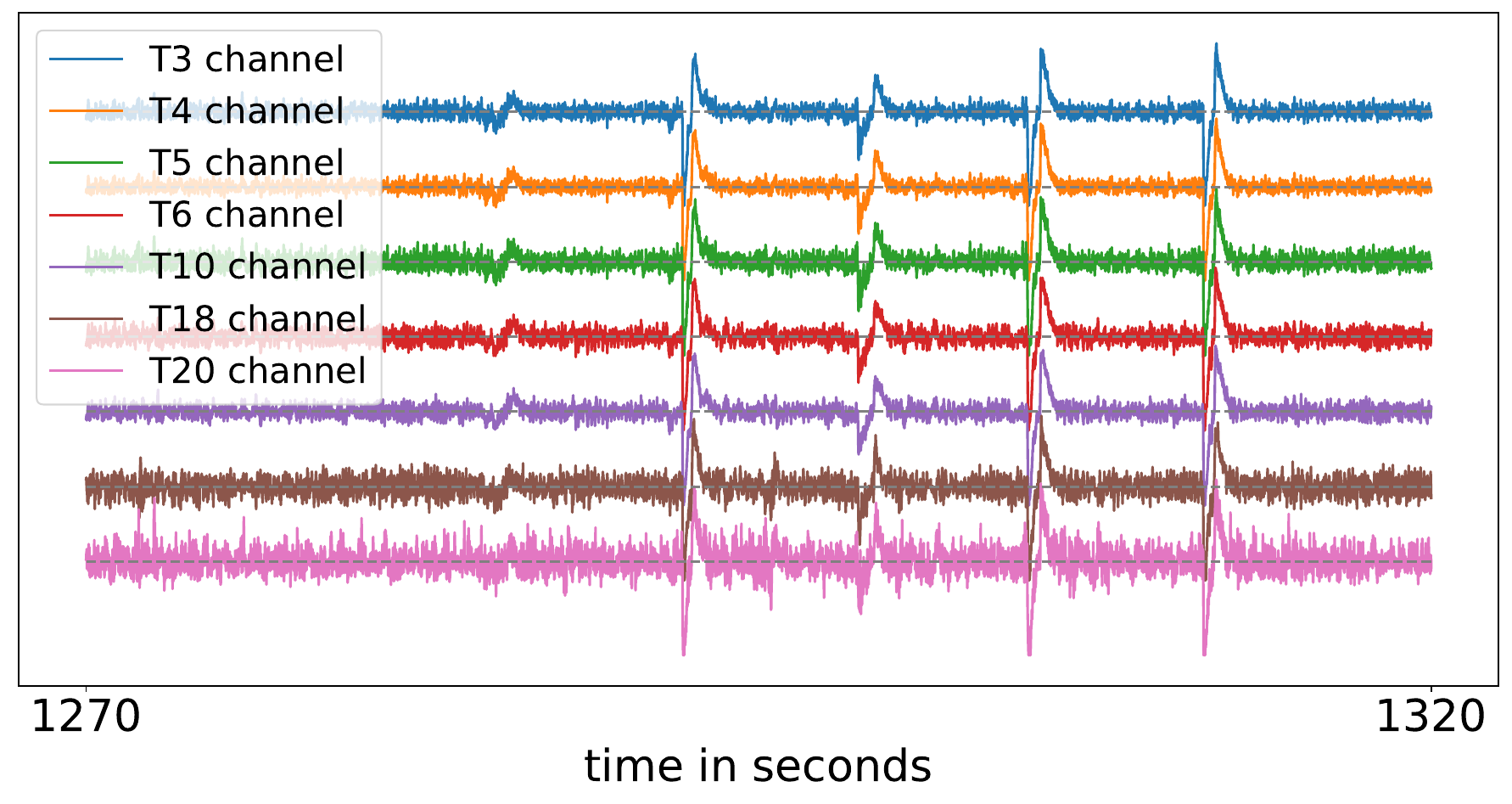}
      \caption{LFP recordings from seven tetrodes of the rat.}
      \label{fig:rat_LFP}
    \end{minipage}
\end{figure*}

Many time series data recorded from various research areas 
including finance, econometrics, biology, atmospheric sciences and neuroscience,  exhibit nonstationarity. 
However, these time series data may have second moment structures (variance, cross-covariance, spectral matrix) that may evolve over the duration 
of the observation period. In this paper, we examine local field potential (LFP) signals recorded from the hippocampus of a rat (see Figure \ref{fig:rat_sequence} and \ref{fig:rat_LFP}), which display changes in the second moment as the rat is presented with sequences of odors during the experiment. Our goal here is to develop a computationally efficient approach for estimating rat brain functional connectivity that can \underline{track these changes in real-time}. 

There are a number of available methods for analyzing non-stationary 
time series data. One approach is to model the data 
as a realization of some "locally stationary" process which 
essentially assumes that the second moment structure 
(i.e., the spectral matrix) is approximately constant 
within a narrow time interval. This idea was introduced 
in \cite{PRIESTLEY} and then reformulated  
in \cite{FITTING_TS_TO_NONSTAT_PROC} with a framework 
that admits a sequence of consistent estimators for the time-varying spectral matrix. 
These two approaches use the Fourier complex 
exponentials as the stochastic building blocks for representing non-stationary signals. However, 
there are other building blocks for signal representation 
In \cite{SLEX} and \cite{NASON_WAVELET} the representations use non-decimated wavelets and SLEX (smooth localized complex exponentials) as building blocks. Thus, under 
these stochastic representations, the time-varying auto-spectra and time-varying cross-coherence are 
rigorously defined.

Another approach to modeling non-stationary signals is to use time-domain representations with time-varying coefficients. Vector autoregressive (VAR) models are popular because they are a natural extension of the autoregressive model to the multivariate setting and thus can capture the cross-dependence between different components. Under the VAR model, the conditional mean is explicitly expressed as a linear combination of its own past as well as the past values of the other time series. Thus, 
the VAR model provides a direct mechanism for forecasting future values of one time series based on its own past values as well 
as the past value of others, see \cite{SIMS} and \cite{LUTKEPOHL}. In general, VAR models are extensively used in numerous fields such as economics, weather forecasting and brain imaging, see \cite{TSA_BROCKWELL_DAVIS}, \cite{TSA_SHUMWAY_STOFFER}, \cite{GORROSTIETA}, \cite{STROKE_AND_BRAIN_MOTOR_FUNCTION}, \cite{Balqis_Connectivity}, \cite{HIERARCHICAL_BAYESIAN_MODEL}.

Some of the main qualities that make VAR models attractive to practitioners are mainly their ability to capture contemporaneous and lagged linear relationships between different components of a time series and also their ability to provide analytical definitions of connectivity based on the spectral density information, see  \cite{GRANGER}, \cite{Bacala_Connectivity}, \cite{Wang_Yuxiao_HighDim_Connectivity}, \cite{Ting_CM}. For a technical review of the VAR models see \cite{LUTKEPOHL}. There are some approaches that allow offline estimation of the parameters of high-dimensional time series under the VAR model. For example the use of regularization techniques, such as the $l_1$ regularization (LASSO-type) can ensure sparsity in high-dimensional settings (see \cite{LASSO}, \cite{BASU_MICHAILIDIS} and \cite{HIGH_DIM_BRAIN_SIGNALS}). Other examples leverage the low-rank property of the transition matrix of the VAR model to gain insight on the network of Granger causal interactions between time series components (see \cite{BASU_LI_MICHAILIDIS} and \cite{Ting_CM}).

Over the last two decades, several methods have been proposed for analyzing non-stationary signals including, \cite{NASON_WAVELET}, \cite{SLEX}, \cite{DAHLHAUS_SUBA_RAO_1}, \cite{DAHLHAUS_SUBA_RAO_2}, \cite{ROBERT_KASS_CMU}, \cite{TIM_PARK}, \cite{Balqis_Connectivity}, \cite{ADAPTIVE_BAYESIAN_SPECTRAL_ANALYSIS}, \cite{Ting_CM}, \cite{LINDQUIST_DYNAMIC_FUNCTIONAL_CONNECTIVITY}, \cite{LINGGE_LI}, \cite{RAQUEL_PRADO} and \cite{ADAPTIVE_BAYESIAN_SPECTRAL_ANALYSIS_HIGH_DIM}. However, none of these
methods can track changes in local field potentials in real-time. 
This is a significant limitation since these methods are not able to provide immediate feedback to the experimenter about, for instance, whether or not the stimulus type and intensity are eliciting the anticipated brain response. Hence, the adequacy of the experimental settings can be verified only retrospectively, that is, after the data acquisition process has been completed. Here, we will develop a procedure that has the ability to provide immediate feedback regarding the connectivity. Thus, the experimenter would be able to adapt the stimulus type or intensity in real-time during the experiment, rather than after the fact. This opportunity for real-time manipulation is critical to investigate the dynamic nature of information processing in the brain. Therefore, we envision two main benefits to the development of robust tools for online estimation of spectral power and coherence in electrode arrays.
First, these tools could have an immediate impact on neuroscience by improving the data collection process. This is particularly important for technically challenging and time-consuming projects, like the rat LFP experiment described here, for which data collection on a given day represents the culmination of months of work (i.e., training the animal, building and surgically implanting the microdrive, and slowly driving the electrodes to the target location). In this case, accurate online estimation of LFP power and coherence could help improve electrode positioning and determine whether the neural response to specific stimuli or conditions is of sufficient effect size to warrant beginning data collection.
Second, these tools are also critical to the ongoing development of the next generation of neuroscience experiments, including closed-loop approaches in which experimental manipulations are performed based on the detection of specific activity patterns in real-time. In this case, rapid and robust online estimation of LFP power and coherence is critical to quickly identify key neural states and trigger the corresponding experimental intervention within the necessary time window. These interventions can range from to the manipulation of experimental parameters, such as the precise timing and intensity of stimuli, to the manipulation of neural circuits through electrical or optogenetic stimulation, to measure the effect of disrupting upstream regions on local information processing.

Our goal in this paper is to develop a procedure that is able to describe these changes of connectivity in local field potentials in real-time. These non-stationary local field potentials will be modeled using time-varying vector autoregressive (tv-VAR) processes. 
Under the tv-VAR framework, we will develop an online procedure for estimating the parameters. Consequently, the procedure will also be able to estimate any functional of these parameters including cross-coherence, partial cross-coherence and partial directed 
coherence. In \cite{TVMVAR}, it have been shown that the tv-VAR model has the ability to capture the dynamics in the multivariate brain signals. However, this comes at a heavy computational cost due to the high dimensionality of the parameter space (which is quadratic in the dimension of the multivariate time series).
In this paper we present a novel smooth online parameter estimation (SOPE) procedure using the penalized least squares criterion. Compared to the Kalman filter, which is the industry standard, the proposed SOPE method has the advantage of being computationally very efficient and that it has 
a Bayesian interpretation. In Section 2 we describe the model, introduce the idea behind smooth online parameter estimation (SOPE), and discuss the infill behavior of our estimator as a function of the penalization parameters. In Section 3 we present a brief overview of some classical brain connectivity measures that are derived from the tv-VAR model. In Section 4 we show that SOPE provides similar performance as the Kalman filter, in terms of mean square error (MSE) using simulated data. However, the main advantage of SOPE is the significantly shorter 
computational time. In Section 5 we apply SOPE to LFP data taken from the hippocampus of a rat performing a hippocampus-dependent sequence memory task and the proposed SOPE method shows results consistent with the literature. Finally, in Section 6 we present a brief discussion of our results.

\section{Smooth Online Parameter Estimation (SOPE)}
\subsection{The tv-VAR model} 
Let $X(t) = [X_1(t), ..., X_P(t)]'$ be a $P$ dimensional time series collected from a network with $P$ nodes.
Here, $X(t)$ could represent the observed brain signal (e.g., electroencephalogram or local field potential) over $P$ locations (electrodes) on the scalp, over time points $t=1, 2, \hdots, T$. Thus, the tv-VAR model of order $K$, denoted tv-VAR($K$), is defined as follows:
\begin{align}
    X(t) = \sum_{\ell =1}^K \Phi_{t,\ell}X(t-\ell) + E(t),
\end{align}
\noindent where the set $\{\Phi_{t,\ell}\}_{\ell=1}^K$ represents the tv-VAR parameters at time $t$ ($K, \hspace{.1cm} P \times P$ matrices); $E(t)$ is Gaussian white noise term with zero mean and variance $\Sigma_E$. In the following we will denote the concatenated unknown time varying parameter matrices by $\Phi(t)$, and we will denote the concatenated previous $K$ observations at time $t$ by $U(t)$:
\begin{align}
    \Phi(t) &= [\Phi_{t,1}, \hdots, \Phi_{t,K}], \label{Eq:tv-VAR_parameters}\\
    U(t) &= [X(t-1)', \hdots, X(t-K)']', \label{Eq:past_observations}
\end{align}
Now the model definition simplifies to
\begin{align}
    X(t) = \Phi(t)U(t) + E(t)
\end{align}
and the parameter $\Phi(t)$ will be estimated in real-time. 

\subsection{Recursive least squares estimators}

The least squares method is a very powerful and fundamental estimation procedure in statistics which dates back to the 
work of Carl Friedrich Gauss and Adrien-Marie Legendre, see \cite{LS_STIGLER_GAUSS}. For almost 150 years since its inception, the least squares estimation procedure did not attract much attention until the early 
1940's when  
\cite{KOLMOGOROV} and \cite{NORBERT_WIENER} independently introduced the field of linear filtering from the stochastic processes point of view. It was only near a decade later   that the efficient matrix formulation of the recursive least squares (RLS) was proposed in \cite{PLACKETT_RLS}. In \cite{KALMAN}, the following major development to the field of linear filtering was proposed using the state space formulation of the problem. More details on the historical development of the RLS/linear filtering can be found in \cite{SORENSON_FROM_GAUSS_TO_KALMAN} and \cite{PETER_YOUNG_RLS}. Dahlhaus and Subba Rao investigated online and recursive inference for time varying ARCH models in \cite{DAHLHAUS_SUBA_RAO_1} and \cite{DAHLHAUS_SUBA_RAO_2}. However, their model can only handle a one dimensional time series, which is not suitable for multivariate time series, as it is often the case for local field potentials.

There are two main paradigms under the general RLS framework. The first is the classical RLS which corresponds to the stationary case where the parameters are static. The second is the weighted RLS which corresponds to the non-stationary case where parameters are evolving over time. Naturally, our interest lies in the second case since the brain functional signals are dynamic. There are two general estimation approaches to weighted RLS: sliding window and exponentially weighted RLS.
To estimate the unknown parameters at a fixed time point, the sliding window recursive least squares (SWRLS) uses a fixed number of observations around this particular time point. By sliding the window across time, this results in time varying parameter estimates where the estimator only borrows information from time-localized observations. However, this approach suffers from high sensitivity to noise in the observations. A solution to this problem is to take a larger window (that is, include more observations), but these results in a more biased estimator due to the bias-variance trade-off. Similarly, the exponentially weighted least squares (EWRLS) uses exponentially decaying weights to estimate the time varying parameters. Therefore, this approach uses all past observations but gives a lower weight to temporally distant observations in order to limit their contribution to the present estimate. In order to control the bias-variance trade-off, a ''forgetting factor" $\kappa \in (0,1)$ is selected where values of $\kappa$ are close to 0 will give more importance to recent observations and therefore will produce estimators with lower bias but these will have higher variance. Conversely, values of $\kappa$ that are close 1 will result in a more biased estimator but with a smaller variance. There is an extensive literature concerning the choice of the decaying weights for the RLS algorithms depending on the problem at hand. However, the flexibility of such algorithms is limited due to the linear nature of the estimator. The above mentioned approaches try to reduce the variance by increasing the window size but this leads to a more biased estimator. Furthermore, those approaches do not leverage potential prior information concerning the smoothness of the estimated curves. In contrast, the proposed SOPE method has the ability to produce estimates with lower variance through the roughness penalty and at the same time preserve the local structure so that the bias is controlled.

\subsection{Kalman filter}

The Kalman filter (KF) is also a recursive algorithm. 
In fact, under simplistic conditions, the RLS is a 
special case of the KF . To estimate $\{\Phi(t)\}$ using the KF, we need to formulate the problem under the framework of state-space models. First, vectorize the parameters (\ref{Eq:vectorized_parameters}) and define observation matrix (\ref{Eq:observation_matrix}):
\begin{align}
    a(t) &= vec(\Phi(t)'), \label{Eq:vectorized_parameters} \\
    C(t) &= I_P \otimes vec([X'(t-1),...,X'(t-K)]). \label{Eq:observation_matrix}
\end{align}
Using the above notation, the formal model is defined in terms of the state transition and observation model, respectively, as follows:
\begin{eqnarray}
    a(t) & = & a(t-1) + w(t), \label{Eq:state}\\
    X(t) & = & C(t) a(t) + v(t), \label{Eq:observation}
\end{eqnarray}
where $w(t)$ and $v(t)$ are the transition noise and observation noise, respectively. The dynamics of the parameters are governed by Equation~\ref{Eq:state}. Under appropriate assumptions (see \cite{KALMAN}), the KF can deliver optimal results in terms of minimal variance of the error. However, the problem at hand is more complicated because the optimal parameters of the filter are unknown and will have to be estimated along with the tv-VAR parameters. Hence, the KF will not necessarily be optimal in the sense of minimal error variance. Indeed, the automatic selection of the filter variance matrices is possible, but that is another problem on its own. For more details on adaptive filtering see \cite{ADAPTIVE_FILTERING_1}, \cite{ADAPTIVE_FILTERING_2}, \cite{ADAPTIVE_FILTERING_3}.

There are two main limitations of the Kalman filter as a tool for practical online estimation. The primary problem is the high level of uncertainty (variance) of the parameter estimators. Consequently, since connectivity measures (e.g., coherence, partial directed coherence) are highly non-linear functions of the tv-VAR parameters, small perturbations in the tv-VAR estimates can lead to substantial changes in the connectivity estimates and hence an even higher level of uncertainty in the connectivity estimators. The other problem of the KF is its prohibitively high computational cost which requires manipulation of a covariance matrix of the state vector. 
Since the state vector lives in a $KP^2$-dimensional space, its covariance matrix will be very expensive to compute and to store in memory when the dimension $P$ of the problem 
is high, which is the case with EEG and LFP  data where $P$ is in the range $20 \sim 256$.

\subsection{Proposed Method: SOPE} \label{sbsec:SOPE_method}

We now propose the SOPE method which leverages the local stationarity assumption to provide real-time estimates of the tv-VAR model parameters. Under local stationarity, 
there is some smoothness assumption on the time-varying parameters. Thus, in our approach we consider the parameters to be differentiable to some given order with respect to rescaled time.

Under Gaussianity ($E(t) \overset{iid}{\sim} \mathcal{N}(0, \Sigma_E)$), the generalized least squares approach (GLS) is equivalent to maximizing the conditional likelihood. Using the notation in Equations \ref{Eq:tv-VAR_parameters} and \ref{Eq:past_observations} we obtain the the following:
\begin{align}
    X(t) \big| \Phi(t), U(t) \sim \mathcal{N}(\Phi(t)U(t), \Sigma_E) \\ \implies \widehat{\Phi}(t) = \underset{b\in \mathbb{R}^{P\times KP}}{\arg \min} \big|\big| X(t) - bU(t) \big|\big|_{\Sigma_E^{-1}}^2. \notag
\end{align}
In this particular problem the MLE will not be an optimal choice for obvious reasons which we state here for completeness. First, the problem is ill posed since $U(t)U(t)'$ is singular, there is only have one observation $X(t)$ at time $t$ to estimate the  tv-VAR parameters $\Phi(t)$. Second, even with regularization the estimator will be biased (shrinkage to zero) and will have a high variance due to the fact that only data from a single time point is used for estimation. Therefore, it will be necessary to borrow information from neighboring time points. However, since our goal is to develop an online estimation method, this neighborhood will consist only of {\it past} observations.

Our proposed method is based on the framework of a locally stationary process, which implicitly assumes some smoothness of the time-varying parameters, and hence the "immediate" past observations must contain relevant information for the current observations. Moreover, one important assumption here is that, for each time $t$, the true physiological signal-generating process can be approximated by a VAR model with parameters $\Phi(t)$ that change smoothly with respect to (rescaled) time. Therefore, a meaningful approach is to obtain an estimator that is reasonably smooth and at the same time provides a good fit based on least squares or some objective criterion (such as the likelihood function). This can be formalized in terms of the penalized least squares criterion below: \begin{align}
    \widehat{\Phi}(t) = \underset{b\in \mathbb{R}^{P\times KP}}{\arg \min} \big|\big| X(t) - bU(t) \big|\big|_{\Sigma_E^{-1}}^2 + \lambda P(b), \label{Eq:penalized_least_squares}
\end{align}
where $\lambda>0$ is a regularization parameter and $P(b)$ denotes a penalization term that controls the smoothness of the estimated function $\widehat{\Phi}(t)$. In this paper, we will investigate different forms of the penalty term $P(b)$ including the Frobenius norm (note that the Frobenius norm of a matrix $A$ is the square root of the sum of its elements squared, i.e., $||A||_F=\sqrt{tr(A'A)}$) of the first order difference with previous estimates or the second order difference or a combination of the two:
\begin{align}
    P_1(b) &= \big|\big| b - \widehat{\Phi}(t-1) \big|\big|_F^2, \label{Eq:first_order_penalization} \\
    P_2(b) &= \big|\big| b - 2\widehat{\Phi}(t-1) + \widehat{\Phi}(t-2) \big|\big|_F^2, \label{Eq:second_order_penalization} \\
    P_3(b) &= \big|\big| b - \big[\widehat{\Phi}(t-1) + \beta \big(\widehat{\Phi}(t-1) -\widehat{\Phi}(t-2)\big)\big] \big|\big|_F^2. \label{Eq:both_orders_penalization}
\end{align}
The penalty function $P_1(b)$ in Equation \ref{Eq:first_order_penalization} corresponds to the first order difference penalization; $P_2(b)$ in Equation \ref{Eq:second_order_penalization} corresponds to the second order difference penalization. Clearly, $P_3(b)$ in Equation \ref{Eq:both_orders_penalization} is a combination of the previous two penalty functions with penalization parameters $\lambda_1$ and $\lambda_2$, where $\lambda = \lambda_1 + \lambda_2$ and $\beta = \frac{\lambda_2}{\lambda_1 + \lambda_2}$. For the following derivations, we will assume that $\Sigma_E = I$. This assumption will greatly simplify computations without unnecessarily limiting the flexibility of the tv-VAR model for capturing the connectivity structure of the brain network. Note that $\Sigma_E = I$ does not imply that the components of the multivariate time series are independent. Instead, this means that the connectivity in the brain network is fully captured by $\Phi(t)$. Moreover, this assumption also allows the computation to be carried out in an online fashion that is computationally robust. The previous problems are all quadratic in $b$. Therefore, minimizing with respect to $b$ leads to the following recursive formulas:
\begin{align}
    \widehat{\Phi}(t) %
    & = \bigl( X(t)U(t)'  + \lambda \widehat{\Phi}(t-1)\bigr)%
    \label{Eq:first_order_solution} \\%
    & \cdot \Big( U(t)U(t)' + \lambda I \Big)^{-1}, %
    \notag\nonumber\\%
    \widehat{\Phi}(t) %
    & = \bigl( X(t)U(t)'  + \lambda \bigl[\widehat{\Phi}(t-1)
    \label{Eq:second_order_solution} \\%
    & \hspace*{2.5mm} + \big(\widehat{\Phi}(t-1) -\widehat{\Phi}(t-2)\big)\bigr] \bigr)%
    \notag\nonumber\\%
    & \cdot \Big( U(t)U(t)' + \lambda I \Big)^{-1},%
    \notag\nonumber\\%
    \widehat{\Phi}(t) %
    & = \bigl( X(t)U(t)'  + \lambda \bigl[\widehat{\Phi}(t-1)  %
    \label{Eq:both_orders_solution} \\
    & \hspace*{2.5mm} + \beta \big(\widehat{\Phi}(t-1) - \widehat{\Phi}(t-2)\big)\bigr] \Bigr) %
    \notag\nonumber\\
    & \cdot \Big( U(t)U(t)' + \lambda I \Big)^{-1} 
    \notag\nonumber.
\end{align}
In order to invert $U(t)U(t)' + \lambda I$ efficiently, it is necessary to use the Sherman-Morrison-Woodbury inversion formula in \cite{INVERSION_FORMULA}.

\subsection{Bayesian interpretation}

The tv-VAR parameters are assumed to vary smoothly over time and therefore they must have a Taylor expansion at every time point $t$. Hence, $\Phi(t) \sim \Phi(t-1) + \dot{\Phi}(t-1)$, where the derivative $\dot{\Phi}(t-1)$ can be approximated by $\Phi(t-1) - \Phi(t-2)$ which leads to $\Phi(t) \sim \Phi(t-1) + \big(\Phi(t-1) - \Phi(t-2)\big)$. Depending on the information available regarding the smoothness of the estimated function and the density of the observations, i.e., the $\Delta t$ between consecutive observations $X(t)$ and $X(t-1)$, this will lead to different levels of confidence in the previous approximation, which translates to $\Phi(t) \sim \Phi(t-1) + \beta\big(\Phi(t-1) - \Phi(t-2)\big)$ with $\beta$ close to one when the function is smooth and when the observations density is very high and close to zero otherwise.

Thus, given estimates $\widehat{\Phi}$ of $\Phi$ at times $t-1$ and $t-2$, a potential prior for the parameters at time $t$ would be $\mathcal{N}(\mu, \Sigma)$, with $\mu = \widehat{\Phi}(t-1) + \beta\big(\widehat{\Phi}(t-1) - \widehat{\Phi}(t-2)\big)$ and $\Sigma = \lambda^{-1} I$, here $\lambda$ is the precision parameter of the prior and it is supposed to represent our confidence on the "location" of the next iterate, and $I$ is a ${KP^2}\times{KP^2}$ identity matrix, the smoother the function the higher the $\lambda$. Of course, moving forward, these parameters need to be vectorized for the interpretation to make sense. Under Gaussianity of the noise, this Bayesian formulation implies the following conditional posterior:
\begin{align}
      f\big(b\big|X(t), U(t)\big) &\propto f\big(X(t)\big|b, U(t) \big) f\big(b\big) \label{Eq:conditional_posterior}
\end{align}
Therefore,  the point estimate $\widehat{\Phi}(t)$ of $\Phi(t)$ at time $t$ is derived to be:
\begin{align}
    \widehat{\Phi}(t) &= \underset{b\in \mathbb{R}^{P\times KP}}{\arg \max} \hspace{1mm} f\big(X(t)\big|b, U(t) \big) f\big(b\big| \mathcal{I}_{t-1} \big), \label{Eq:parameter_MLE_PLSE} \\ &= \underset{b\in \mathbb{R}^{P\times KP}}{\arg \min} \big|\big| X(t) - bU(t) \big|\big|_{\Sigma_E^{-1}}^2 + \lambda P(b). \notag
\end{align}
Note that the maximization problem of the (log-)posterior in Equation \ref{Eq:parameter_MLE_PLSE} is equivalent to the minimization problem stated in Equation \ref{Eq:penalized_least_squares}, with penalties defined in Equations \ref{Eq:first_order_penalization}, \ref{Eq:second_order_penalization} and \ref{Eq:both_orders_penalization} corresponding to different prior choices. A choice of $\beta=0$ is equivalent to the choice of a Gaussian prior that assumes a constant function. This is consistent with penalizing the gradient since this penalization will have as a result the vanishing of the estimate's gradient. Moreover, taking $\beta=1$ is equivalent to having a Gaussian prior that assumes a constant gradient/linear function of time. This is consistent with penalizing the curvature since this penalization produces a vanishing second derivative. Note that it is possible to pursue higher order penalties, one could extend this reasoning indefinitely to chose the corresponding prior that assumes some degree of smoothness.

Given the penalization parameters $\alpha$, $\beta$ and the time series $X(t)$ the following recursive algorithm (SOPE) provides online estimates of the parameters:
\begin{algorithm}[H]
  \caption{Smooth Online Parameter Estimation for tv-VAR models}
  \begin{algorithmic}[1]
    \Procedure{GetSmoothEstimates}{$X(1),\hdots, X(T)$}
    \State Initialize:
    \State $\widehat{\Phi}(K-1)$ = Least Squares
    \State $\widehat{\Phi}(K)$ \hspace{.7cm}= $\widehat{\Phi}(K-1)$
    \State $\alpha \in (0, \infty)$, $\beta \in [0,\hdots, 1)$
    \For{$t=K+1,\hdots,T$}
    \State $U(t) = \big[X(t-1)', \hdots, X(t-K)'\big]'$
    \State $A = X(t)U(t)'  + \lambda \Big[\widehat{\Phi}(t-1) + \beta \big(\widehat{\Phi}(t-1) - \widehat{\Phi}(t-2)\big)\Big]$
    \State $B = \big(U(t)U(t)' + \lambda I\big)^{-1}$
    \State $\widehat{\Phi}(t) = A B$
    \EndFor
    \EndProcedure
  \end{algorithmic}
\end{algorithm}

\subsection{Infill asymptotics}
\label{sbsec:infill_asymptotics}

Leaving the world of stationarity comes with many complications. Classical asymptotic results do not necessarily apply here, since the distant future or the distant past might not provide relevant information about the present due to the nonstationarity. Therefore, by considering locally stationary processes, one should look into infill asymptotics on the parameter $\Phi(t)$. Following \cite{LOCAL_STATIONARITY}, we analyze the behavior of our estimator using infill asymptotics by rescaling the time index of the tv-VAR parameter to the unit interval $[0,1]$. Define $u = \frac{t}{T}$, then as $T\to \infty$, the range of $u$ will become dense in $[0,1]$.

Asymptotically, when $T \to \infty$, the (rescaled) time between consecutive observations becomes infinitesimal, i.e., $h = \frac{1}{T} \to dt$. Furthermore, if the penalization coefficients are selected correctly, the finite difference penalization terms will converge to the derivatives, see Appendix \ref{sbsec:infill_asymptotics}. Moreover, minimizing the problem globally ($\forall t \in [1,\hdots,T]$) with penalty terms as defined in Equations \ref{Eq:first_order_penalization} and \ref{Eq:second_order_penalization} will asymptotically result into the following calculus of variation problems:
\begin{align}
    &\begin{cases} \label{Eq:first_order_calculus_var}
        \mathcal{L}(u, b, \dot{b}) &= \big|\big| X(\lfloor uT \rfloor) - b(u)U(\lfloor uT \rfloor)\big|\big|_2^2 + \\ & \qquad c_1 \big|\big| \dot{b}(u) \big|\big|_F^2 \\
        \mathcal{J}[b] &= \int_{0}^{1} \mathcal{L}(u, b, \dot{b}) du\\
        \widehat{\Phi} &= \underset{b\in B}{\arg \min } \hspace{.1cm} \mathcal{J}[b]
    \end{cases}, \\
    &\begin{cases} \label{Eq:second_order_calculus_var}
        \mathcal{L}(u, b, \dot{b}, \ddot{b}) &= \big|\big| X(\lfloor uT \rfloor) - b(u)U(\lfloor uT \rfloor) \big|\big|_2^2 + \\ & \qquad c_2 \big|\big| \ddot{b}(u) \big|\big|_F^2 \\
        \mathcal{J}[b] &= \int_{0}^{1} \mathcal{L}(u, b, \dot{b}, \ddot{b}) du\\
        \widehat{\Phi} &= \underset{b\in B}{\arg \min} \hspace{.1cm} \mathcal{J}[b]
    \end{cases},
\end{align}
where the first equations represent the Lagrangian which is the least squares term (or the likelihood of the observations) plus a penalty term for the roughness of the function; the second equations represent the functionals to be minimized over $B$ (class of smooth enough functions). To solve such problems, we consider the necessary condition for optimality, also known as Euler-Lagrange equations (see Appendix \ref{sbsec:infill_asymptotics} for more details and \cite{CALCULUS_VAR} for more background on Euler-Lagrange equations and calculus of variation in general):
\begin{align}
    \ddot{\widehat{\Phi}}(u) &= \frac{1}{c_1} \nabla_{\widehat{\Phi}} ||X(\lfloor uT \rfloor) - \widehat{\Phi}U(\lfloor uT \rfloor)||_2^2, \label{Eq:first_order_euler_lagrange} \\
    \dddot{\widehat{\Phi}}(u) &= -\frac{1}{c_2} \nabla_{\widehat{\Phi}} ||X(\lfloor uT \rfloor) - \widehat{\Phi}U(\lfloor uT \rfloor)||_2^2. \label{Eq:second_order_euler_lagrange}
\end{align}
If the derivative is penalized with $c_1$ being small (asymptotically "weak" penalization) then the solution boils down to the least squares estimator. However, if $c_1$ is large, the second derivative vanishes which implies that the solution is affine:
\begin{align*}
    \begin{cases}
        c_1 \to 0 &\implies \forall u, \nabla_{\widehat{\Phi}} ||X(\lfloor uT \rfloor) - \widehat{\Phi}(u)U(\lfloor uT \rfloor)||_2^2 \to 0 \\
        c_1 \to \infty &\implies \forall u, \ddot{\widehat{\Phi}}(u) \to 0,
    \end{cases}
\end{align*}
Similarly, if we penalize the second derivative with $c_2$ being small (asymptotically "weak" penalization) the solution boils down to the least squares estimator. However, if $c_2$ is large, the third derivative vanishes which means that the solution is quadratic in time:
\begin{align*}
    \begin{cases}
        c_2 \to 0 &\implies \forall u, \nabla_{\widehat{\Phi}} ||X(\lfloor uT \rfloor) - \widehat{\Phi}(u)U(\lfloor uT \rfloor)||_2^2 \to 0 \\
        c_2 \to \infty &\implies \forall u, \dddot{\widehat{\Phi}}(u) \to 0.
    \end{cases}
\end{align*}
Therefore, based on the choice of the penalty paramters $c_1$ and $c_2$, the SOPE method will produce an estimator that is an intermediate between the LSE (or MLE) on one side and the best linear (or parabolic) curve on the other side depending on the exact regularization that is used.


Note that if $\Sigma_E$ is unknown, it needs to be estimated online. Therefore, the previous results will not be valid anymore. However, if the covariance estimates improve as more data is observed, then previous results will hold after some burn in period that is necessary for getting a good estimate for $\Sigma_E$, see Appendix \ref{sbsec:generalized_sope} for the generalized SOPE algorithm.

\subsection{SOPE and the Kalman Smoother}

In addition to the previous two subsections, it is possible to make a connection between smooth online parameter estimation as described in Equation \ref{Eq:both_orders_penalization} and state space models that bridges the gap between the Bayesian interpretation and the asymptotic behaviour of the obtained estimator.

On the one hand the overall SOPE estimator could be summarized as the following minimization problem
\begin{align}
    \widehat{\Phi}  &= \underset{b_K, \hdots, b_T}{\arg \min} \hspace{.1cm} \sum_{t=K}^T \big|\big| X(t) - b(t)U(t) \big|\big|_2^2 + \lambda \big|\big| \ddot{b}(t) \big|\big|_F^2
    \label{Eq:overall_SOPE}
\end{align}
with respect to the time varying VAR parameters. Where $\lambda$ controls the smoothness of the estimates, the larger the $\lambda$ the smoother the estimates. If $\lambda=0$, the minimizer is the OLS estimator, which leads to non smooth estimates. If $\lambda=\infty$, then the only acceptable solution is one that vanishes the second derivative, which in turn leads to an estimator that is linear in time as presented in Section \ref{sbsec:infill_asymptotics}.

On the other hand assuming the following model
\begin{align}
    X(t) = \Phi(t)U(t) + v(t) \text{ and } \ddot{\Phi}(t) = w(t) \label{def:state_space_model}
\end{align}
where $v(t)$ and $w(t)$ are independent white noise processes with $Var(v(t)) = \sigma_v^2I$ and $Var(w(t)) = \sigma_w^2I$. One can show that the model defined in \ref{def:state_space_model} leads to a state space model, where minimizing the log likelihood with respect to the states is equivalent to maximizing the complete data likelihood which is equivalent to the minimization problem in Equation \ref{Eq:overall_SOPE} with $\lambda=\frac{\sigma_v^2}{\sigma_w^2}$. Thus, the variance $\sigma_w^2$ controls the smoothness of the estimates, the smaller $\sigma_w^2$ the larger $\lambda$ and the smoother $\widehat{\Phi}(t)$. This connection between the SOPE and the Kalman smoother is somehow very similar to the connection between smoothing splines and state space models. Instead of smoothing the data itself (i.e., obtaining $\widehat{\mu}(t)$ from $y_t = \mu(t) + \epsilon(t)$) we aim to smooth the time varying parameters. Refer  to \cite{TSA_SHUMWAY_STOFFER} for more details on how to build the state space model from Equation \ref{def:state_space_model} and to see more details on the connection between smoothing splines and the Kalman smoother in general.

\section{Online Estimates of Brain Connectivity Measures} \label{sbsec:connectivity_measures}

For centuries, scientists have been interested in localizing brain functions (e.g., \cite{GALL_PHRENOLOGICAL_MOVEMENT}). In recent decades, considerable progress has been made bringing to light the anatomical and structural map of the brain. 
However, understanding how information is processed and integrated across brain regions is a much more difficult challenge because it involves dynamics and the notion of causality (see \cite{KARL_FRISTON} and \cite{STATISTICAL_ANALYSIS_BRAIN_DATA}).

Functional connectivity is often defined as the statistical dependence between distant populations/groups of neurons. Depending on the data modality (in particular, the sampling rate) it is usually assessed using cross-correlation (for functional magnetic resonance imaging data) or cross-coherence (for electroencephalograms or local field potentials). Over the last twenty years there have been an increasing trend in the literature related to the assessment of effective connectivity, which can be understood as the influence of a group of neurons over another, see \cite{DYNAMIC_CONNECTIVITY_REGRESSION}, \cite{DYNAMIC_FCONNECTIVITY}, \cite{DYNAMIC_BRAIN_PROCESSES}, \cite{CROSS_NEURONAL_INTERACTIONS}, \cite{TRIAL_BRAIN_SIGNALS_MODELS}, \cite{DYNAMIC_COMMUNITIES}. Consequently, effective connectivity is a measure of causal influence/information flow between two regions. Baccala and Sameshima developed the concept of partial directed coherence (PDC) between groups of neurons which measures the direction of information flow inside the brain structure, see \cite{HARASHIMA} and \cite{Bacala_Connectivity}.

In this paper, we discuss three measures of connectivity, coherence, partial coherence and partial directed coherence. However, in our data analysis, we will only use coherence and partial directed coherence. Coherence is the relative synchrony between a pair of signals. Coherence at some pre-specified frequency band is the squared cross-correlation between a pair of filtered signals (whose power are each concentrated at the specific band). This intuitive exposition of coherence is given in \cite{OMBAO_BELLEGAN}. Partial coherence between two signals can be defined as the conditional coherence when all other signals have been observed. In other words partial coherence is the remaining coherence that cannot be explained by all other signals. It is characterized via the inverse of the spectral matrix. To handle estimation problems when the spectral matrix has a poor condition number, shrinkage estimators are proposed in \cite{SHRINKAGE_FUNCTIONAL_CONNECTIVITY} and \cite{GENERALIZED_SHRINKAGE_FUNCTIONAL_CONNECTIVITY}. Partial directed coherence (PDC) can be defined as the conditional Granger causality from one signal to another normalized by the total causal outflow from the first signal and is developed under the context of a vector autoregressive model. PDC was introduced in the seminal papers of \cite{HARASHIMA} and \cite{Bacala_Connectivity}, and was later used by \cite{HIGH_DIM_BRAIN_SIGNALS} and \cite{HIERARCHICAL_BAYESIAN_MODEL} in the context of modeling high dimensional signals. 

In the following, we provide the parametric definitions of these time-varying spectral measures of connectivity in the context of a tv-VAR model. First, we define
the following time varying quantities of interest at frequency $\omega$ with sampling frequency $\omega_s$:
\begin{align}
    \Phi(t, \omega) &= I_P - \sum_{\ell=1}^K \Phi_{t,\ell} \exp{(-\sqrt{-1}2\pi \ell \omega/\omega_s)}, \label{Eq:fourier_parameters} \\
    H(t, \omega) &= \Phi(t, \omega)^{-1}, \label{Eq:transfer_matrix}\\
    S(t, \omega) &= H(t, \omega) \Sigma_E H(t, \omega)^*, \label{Eq:spectral_matrix}
\end{align}
$S(t, \omega)$ is called the spectral matrix of the time series, $S_{i,j}(t, \omega)$ represents the cross spectrum between tetrode $i$ and tetrode $j$ at frequency $\omega$ and time $t$, $H(t, \omega)$ is the time-varying and frequency-specific transfer function matrix.
Using the previous notation in Equations \ref{Eq:fourier_parameters}, \ref{Eq:transfer_matrix} and \ref{Eq:spectral_matrix} we can provide the parametric definitions of the spectral measures of dependence defined above between tetrode $i$ and tetrode $j$ at frequency $\omega$ and time $t$:
\begin{align}
    \rho^2_{i,j}(t, \omega) &= \frac{|S_{i,j}(t, \omega)|^2}{S_{i,i}(t, \omega)S_{j,j}(t, \omega)}, \label{Eq:coherence}\\
    \pi_{i, j}(t, \omega) &= \frac{\big|\Phi_{i,j}(t, \omega)\big|}{\sqrt{\sum_{k=1}^P \big|\Phi_{k,j}(t, \omega)\big|^2}},\label{Eq:partial_directed_coherence}
\end{align}
The previous Equations \ref{Eq:coherence} and \ref{Eq:partial_directed_coherence} represent respectively coherence and partial directed coherence. To estimate these spectral connectivity measures at time $t$, first a tv-VAR model is used to estimate in real-time the time-varying parameters, then it suffices to plug in the above formulas to get the spectral connectivity measures. Therefore, online estimates of the tv-VAR model parameters will naturally provide online plugin estimator for spectral connectivity:
\begin{align}
    \widehat{\rho}^2_{i,j}(t, \omega) &= \frac{|\widehat{S}_{i,j}(t, \omega)|^2}{\widehat{S}_{i,i}(t, \omega)\widehat{S}_{j,j}(t, \omega)}, \label{Eq:estimated_coherence} \\
    \widehat{\pi}_{i, j}(t, \omega) &= \frac{\big|\widehat{\Phi}_{i,j}(t, \omega)\big|}{\sqrt{\sum_{k=1}^P \big|\widehat{\Phi}_{k,j}(t, \omega)\big|^2}}, \label{Eq:estimated_partial_directed_coherence}
\end{align}

\section{Simulation Studies}

We now investigate the ability of the SOPE method to estimate in real-time the time-varying parameters. The key metrics are computational time and also mean-squared error of the connectivity measures.

\subsection{Computational time}

As it was mentioned in the previous sections, when compared to the KF, the proposed SOPE method has the advantage of being computationally faster, because it does not need to keep track of a covariance matrix of the vectorized parameters. The average execution time per iteration (in milliseconds) to estimate the tv-VAR parameters is reported in the tables \ref{Table:KFTime} and \ref{Table:SOPETime}.
\def\ttitle#1{{\centering\bfseries #1\\}}
\begin{table*}[ht]
    \begin{center}
        \ttitle{Execution time per iteration for Kalman filter}
        \vspace{1mm}
        \begin{tabular}{l|r|r|r|r|r|r|}
        \cline{2-7}
                                  & $P=2$  & $P=5$  & $P=10$  & $P=15$   & $P=20$   & $P=25$    \\ \hline
        \multicolumn{1}{|l|}{$K=1$} & 0.24 & 0.14 & 0.31  & 1.61   & 7.89   & 28.58   \\ \hline
        \multicolumn{1}{|l|}{$K=3$} & 0.11 & 0.25 & 3.60  & 27.35  & 116.66 & 434.45  \\ \hline
        \multicolumn{1}{|l|}{$K=5$} & 0.09 & 0.38 & 14.10 & 102.29 & 517.96 & 1827.42 \\ \hline
        \end{tabular}
    \end{center}
    \caption{All durations are in milliseconds. $P$ represents the dimension and $K$ represents the order of the tv-VAR model.}\label{Table:KFTime}
\end{table*}
\begin{table*}[ht]
    \begin{center}
        \ttitle{Execution time per iteration for SOPE}
        \vspace{1mm}
        \begin{tabular}{l|r|r|r|r|r|r|}
        \cline{2-7}
                                  & $P=20$ & $P=50$ & $P=100$ & $P=150$ & $P=200$ & $P=250$  \\ \hline
        \multicolumn{1}{|l|}{$K=1$} & 0.11 & 0.23 & 0.46  & 1.39  & 2.19  & 3.47   \\ \hline
        \multicolumn{1}{|l|}{$K=3$} & 0.19 & 0.96 & 4.63  & 13.77 & 22.70 & 38.37  \\ \hline
        \multicolumn{1}{|l|}{$K=5$} & 0.34 & 2.04 & 13.79 & 37.96 & 70.34 & 114.85 \\ \hline
        \end{tabular}
    \end{center}
    \caption{All durations are in milliseconds. $P$ represents the dimension and $K$ represents the order of the tv-VAR model.}\label{Table:SOPETime}
\end{table*}

Typically, brain signals such as EEG and LFP have high time resolution (roughly from 256Hz to 2000Hz). Therefore, any method that aims to be applied in real-time needs to have an execution time in the order of a few milliseconds per iteration at most, in order to be able to update the estimates for every observation. Clearly, Tables \ref{Table:KFTime} and \ref{Table:SOPETime} show that the SOPE method is computationally faster than the Kalman filter by orders of magnitude. The SOPE method can handle higher dimensions in the order of 256 (dense EEG setting), as opposed to the Kalman filter that can barely handle small dimensions below 25. Furthermore, the SOPE implementation uses the Python
language which is known to be slow when compared with lower level languages such as C. Therefore, an implementation of the proposed SOPE method in a lower level programming language will be 
useful for applications to even higher dimensions.

\subsection{Parameter estimation}

In the following, we start by providing a simple example to show the quality of the estimates obtained using the SOPE method, we then show a more rigorous comparison between the Kalman filter and the SOPE method in terms of average Sum of Square Errors (SSE).

In the following example, a tv-VAR model with dimension $P=5$ and order $K=1$ is used. The five components are split in two groups of size three and two, the two groups could potentially be represented by individual tv-VAR models of dimension three and two respectively. Every entry of the coefficient matrix $\Phi(t)$ is either a cosine function with random amplitude and random phase shift or a zero function, i.e., $\Phi(t)_{i,j} = A_{i, j} \cos{(\pi \frac{t}{T} + B_{i,j})}$ if $i$ and $j$ are in the same group and $\Phi(t)_{i,j} = 0$ otherwise, where $A_{i, j}$ is the random amplitude and $B_{i, j}$ is the random phase shift. In order to ensure stationarity at every time $t$ the amplitudes are designed to be higher when $i=j$ and smaller when $i\neq j$.

\begin{figure*}
    \centering
    \includegraphics[width=\linewidth]{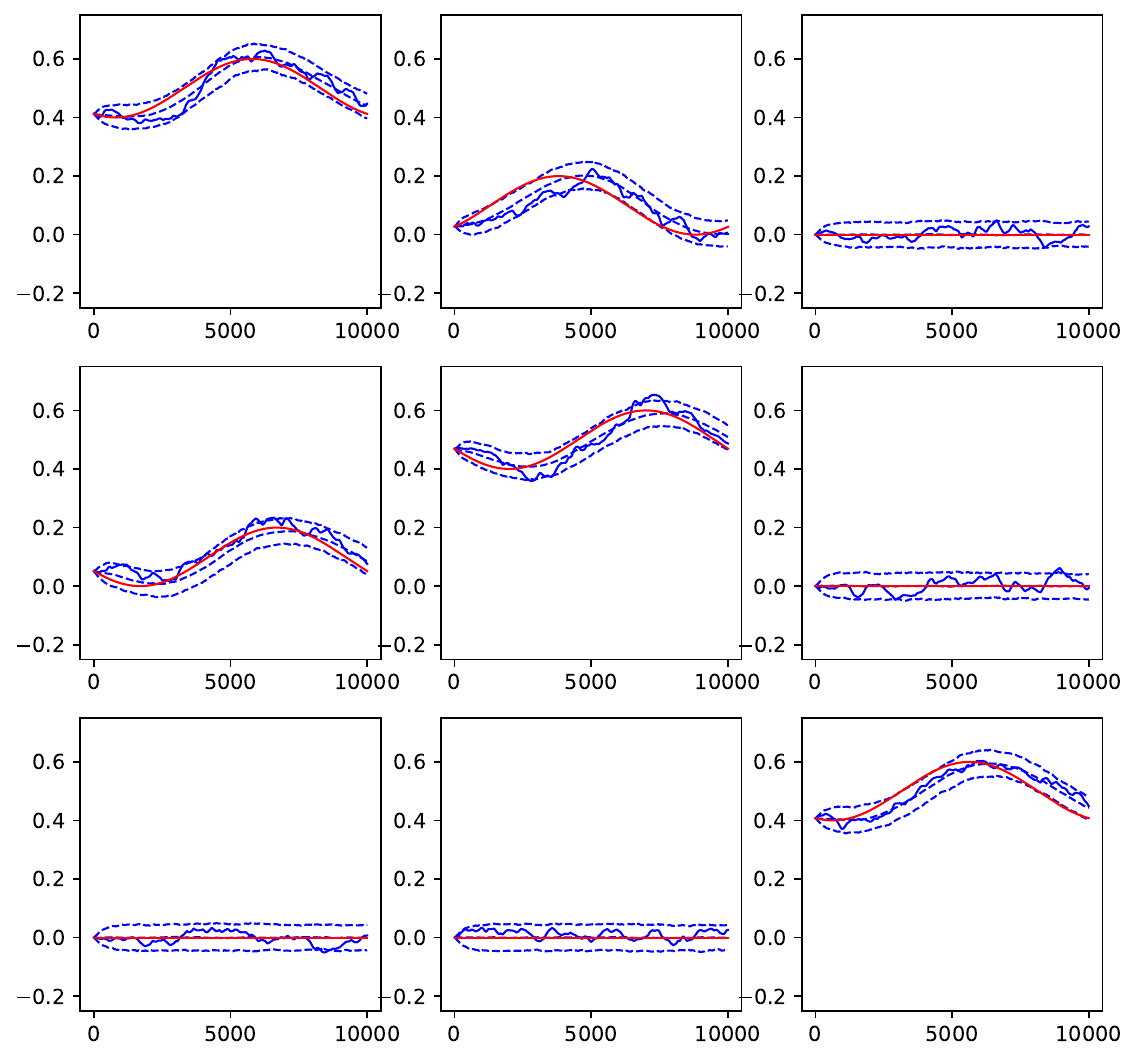}
    \caption{SOPE results for tv-VAR model: the plot in $i^{th}$ row and $j^{th}$ column represent the results for $\big(\Phi_1(t)\big)_{2+i, 2+j}$ (red line) and $\big(\widehat{\Phi}_1(t)\big)_{2+i, 2+j}$ (blue line), the 2.5, 50 and 97.5 percentiles (based on $B=1000$ samples) are represented in dotted blue lines.}
    \label{fig:tv-VAR_PARAMS_SIM}
\end{figure*}

A straightforward implementation of Algorithm $1$ in Section \ref{sbsec:SOPE_method} with identity noise covariance matrix provides the following estimates (only $3\times 3$ submatrix, refer to Appendix Section \ref{subsec:full_param_connectivity_matrices} for the entire parameter matrix) (with $0.025$, $0.5$ and $0.975$ quantiles represented by dotted lines based on 1000 sample estimates), see Figure \ref{fig:tv-VAR_PARAMS_SIM}.

\begin{figure*}
    \centering
    \includegraphics[height=\linewidth, width=.9\linewidth]{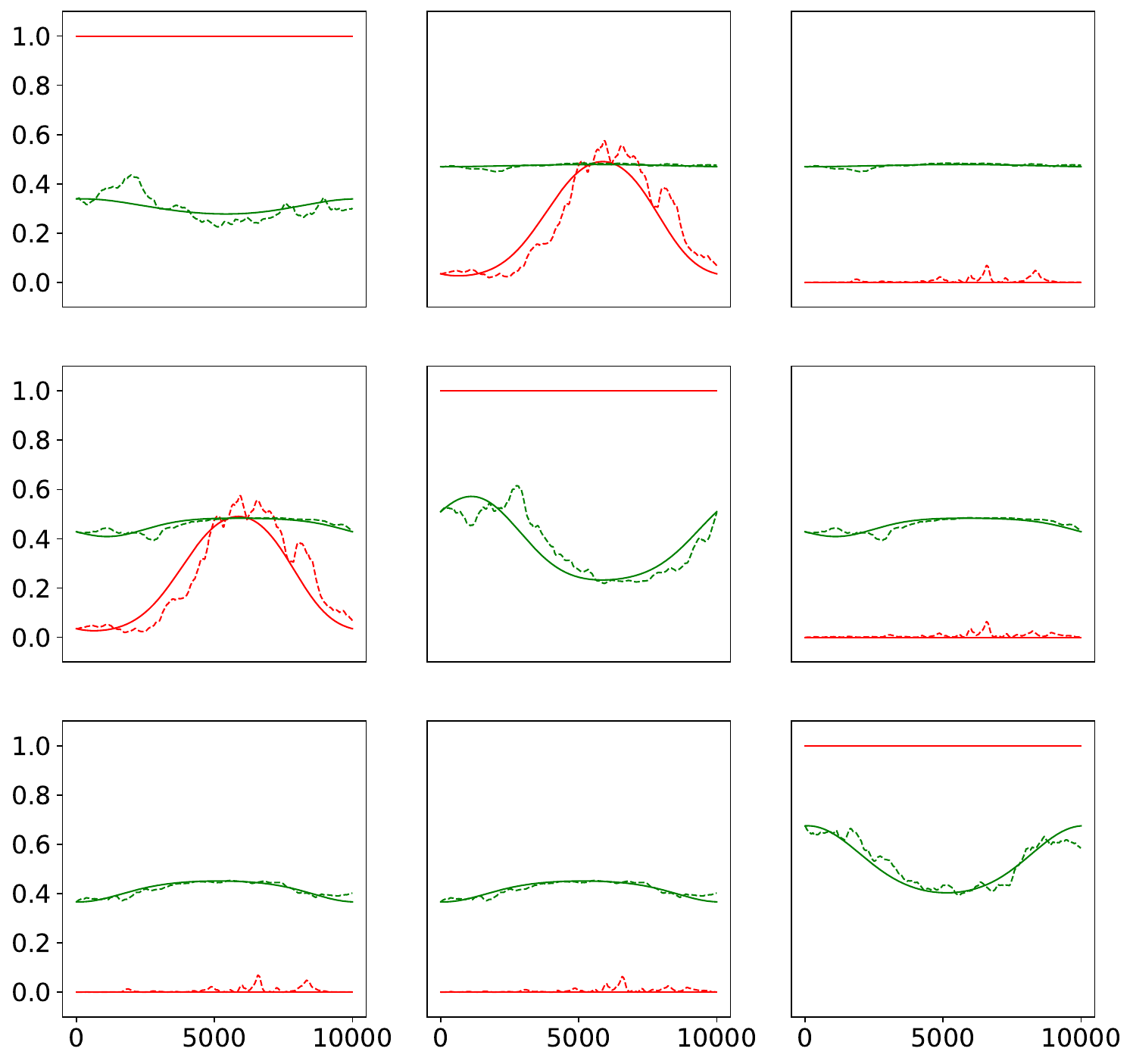}
    \caption{SOPE results for tv-VAR model: the plot in $i^{th}$ row and $j^{th}$ column represent the connectivity estimates between tetrode $2+i$ and $2+j$. Coherence is in red, while PDC is in green, solid lines represent the true quantity and dotted lines represent the estimates (for Delta band).}
    \label{fig:Conectivity_SIM_DELTA_C_PDC}
\end{figure*}

After obtaining the tv-VAR model parameter estimates, it is possible to compute the connectivity measures of interest, for example coherence and PDC as defined in Section 3, refer to Figure \ref{fig:Conectivity_SIM_DELTA_C_PDC} to see the $3\times3$ submatrix estimates and Appendix Section \ref{subsec:full_param_connectivity_matrices} for the full matrix connectivity results. Equations \ref{Eq:estimated_coherence} and \ref{Eq:estimated_partial_directed_coherence} clearly show the nonlinear dependence between the tv-VAR parameter estimates and the measures of connectivity estimates. A small noise on the parameter estimates can result in a large deviation of the estimator of connectivity from the true values. Therefore, having a method that is able to control for the smoothness will result in a significant and practical advantage for practitioners.

The SOPE approach is computationally faster and hence can be applied to higher dimensions. Moreover, it can also provide estimates that compete with the Kalman filter in terms of mean square error. Thus, the SOPE approach does not need to trade shorter computational time with estimation accuracy.

\begin{figure*}[ht]
    \centering
    \includegraphics[width=.9\textwidth]{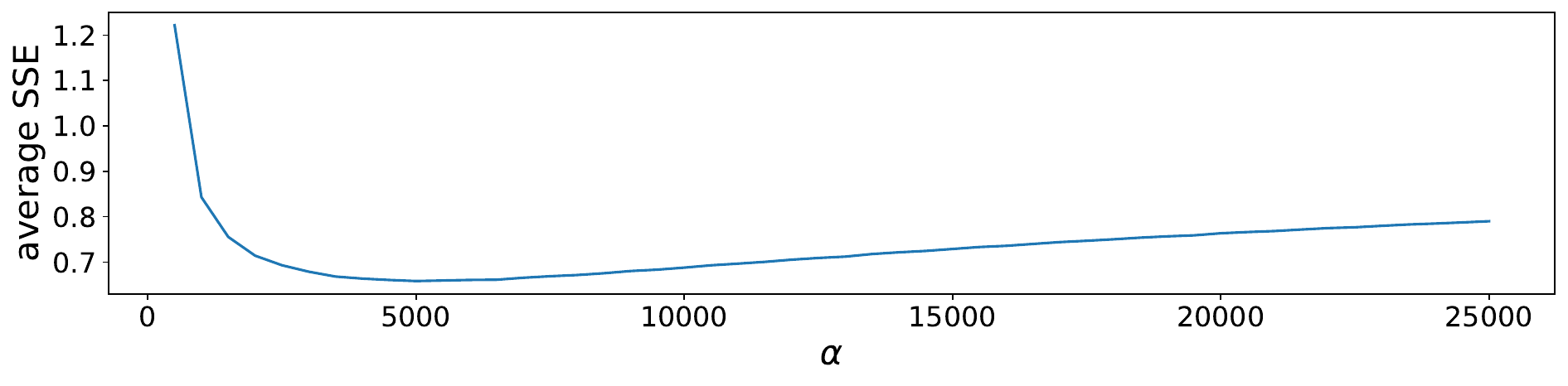}
    \caption{SOPE: average MSE per parameter as a function of $\alpha$ for a tv-VAR model P=3 and K=2.}
    \label{fig:SOPE_SSE_LAM}
\end{figure*}

Both the Kalman filter and the SOPE algorithms have parameters that need to be tuned, the state covariance matrix for the Kalman filter and the regularization parameters $\alpha$ and $\beta$ for the SOPE. In the following simulation, we select the optimal parameters that minimize the MSE for the tv-VAR model parameter estimates (for $P=3$ and $K=2$). For simplicity we take the state covariance as $\Sigma_Q = \sigma^2 I$, experimentally a $0.9$ value for $\beta$ seem to work very well in practice. Figures \ref{fig:KF_SSE_SIG} and \ref{fig:SOPE_SSE_LAM} show the average MSE as a function of the hyperparameters $\sigma$ and $\alpha$:
\begin{figure*}[ht]
    \centering
    \includegraphics[width=.9\textwidth]{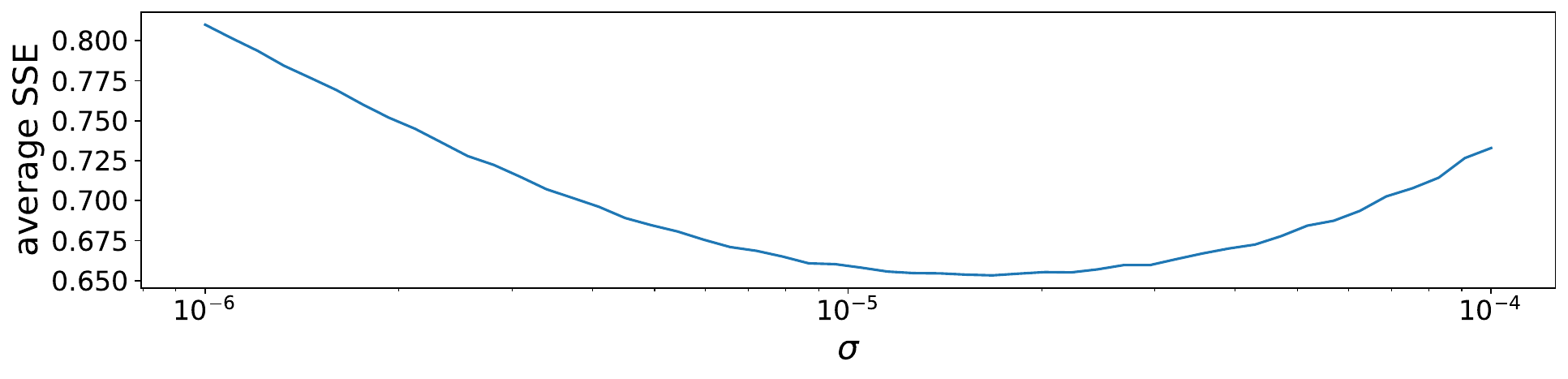}
    \caption{Kalman filter: average MSE per parameter as a function of $\sigma$ for a tv-VAR model P=3 and K=2.}
    \label{fig:KF_SSE_SIG}
\end{figure*}

A few remarks about the results in the previous Figures are as follows. First, both approaches have similar MSE per parameter that is between $0.006$ and $0.007$ with Kalman filter being slightly better. Second, the optimal $\sigma$ is in the order of $10^{-5}$ and the optimal $\alpha$ is in the order of $5000$, which means that the Kalman filter is much more sensitive to the choice of $\sigma$ than the SOPE is to the choice of $\alpha$.

Given the previous simulations that assess the behaviour of the estimators in terms of the MSE with respect to the selection of the hyperparameters, we show here that we can use such knowledge about the hyperparameters and transfer it to another similar but different problem. Thus, using the optimal hyperparameters from the example above, we apply the KF and the SOPE methods to a tv-VAR model with $P=4$ and $K=3$ in order to investigate the parameter estimates for both models. We report bellow the results for the pairs of tetrodes (2, 2) and (4, 2).

\begin{figure*}
    \centering
    \includegraphics[width=.8\textwidth]{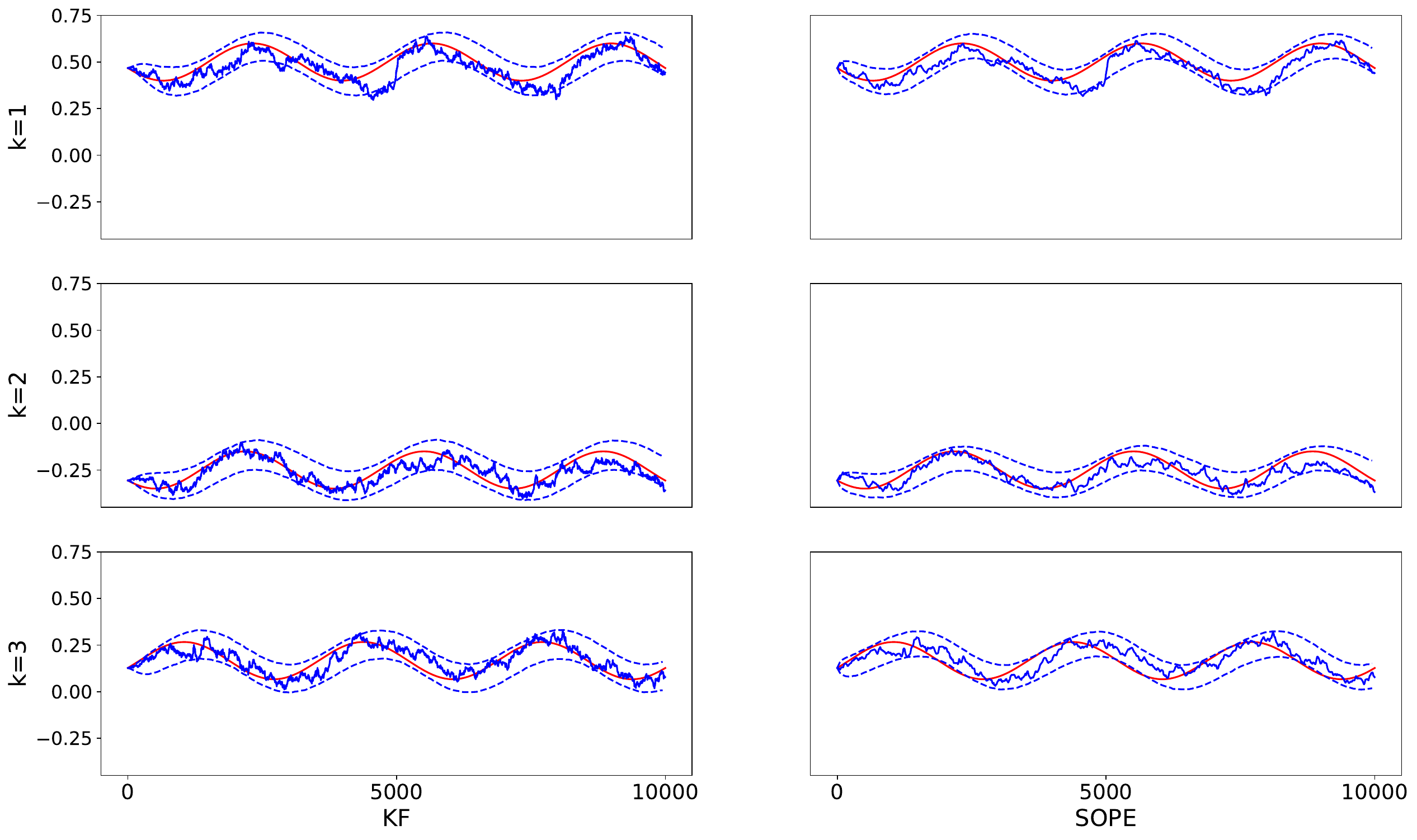}
    \caption{Parameter estimates for a tv-VAR model with $P=4$ and $K=3$, pair of tetrodes $(2, 2)$. True parameters are in solid red lines, estimated parameters are in solid blue lines and $2,5$  and $97.5$ percentiles (based on $5000$ samples) are in doted blue lines. KF on the left column and SOPE on the right column. The results are based on the optimal $\sigma\sim 10^{-5}$ and $\alpha\sim 5000$.}
    \label{fig:KF_SSE}
\end{figure*}

\begin{figure*}[p]
    \centering
    \includegraphics[width=.8\textwidth]{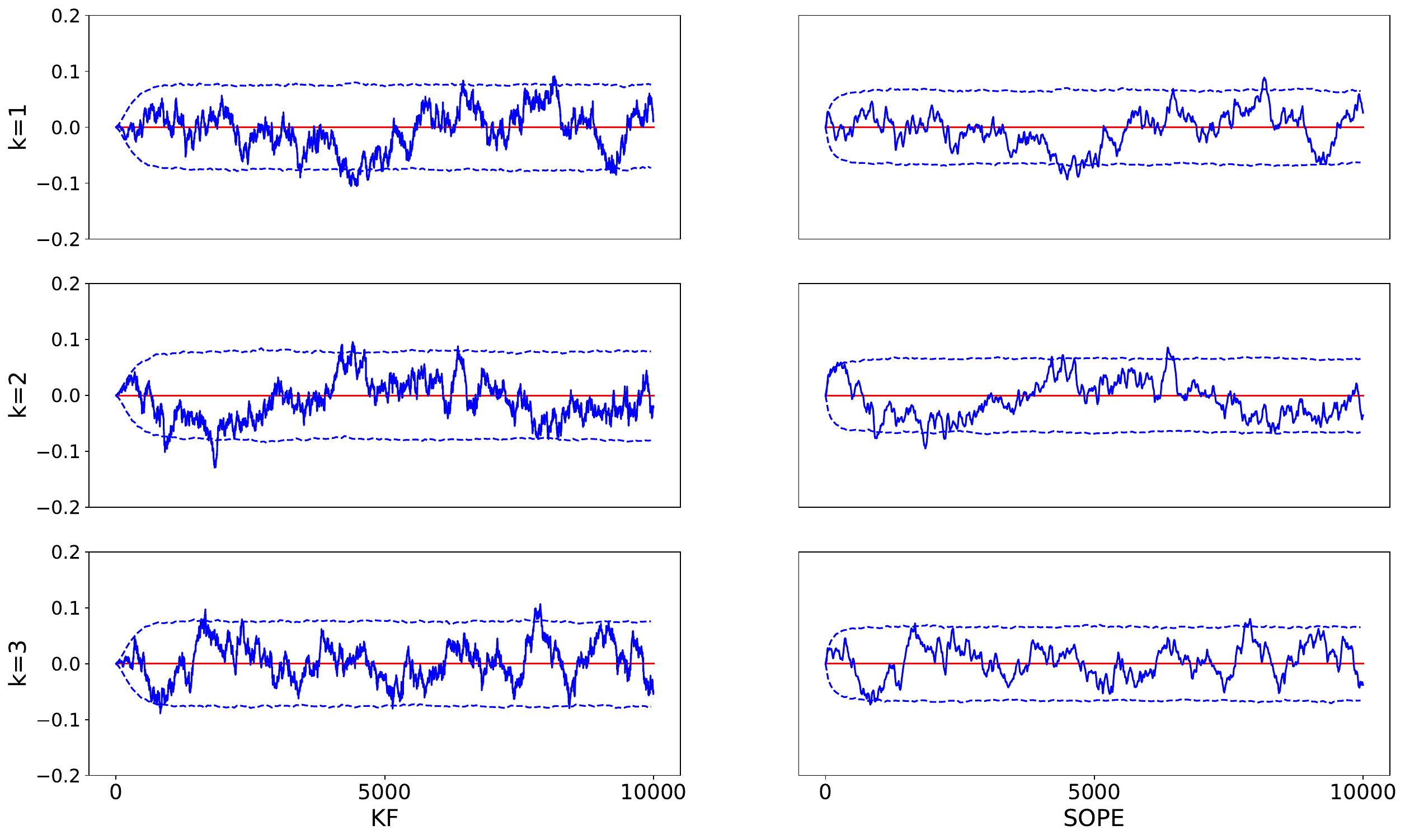}
    \caption{Parameter estimates for a tv-VAR model with $P=4$ and $K=3$, pair of tetrodes $(4, 2)$. True parameters are in solid red lines, estimated parameters are in solid blue lines and $2,5$, $50$ and $97.5$ percentiles (based on $5000$ samples) are in doted blue lines. KF on the left column and SOPE on the right column. The results are based on the optimal $\sigma\sim 10^{-5}$ and $\alpha\sim 5000$.}
    \label{fig:SOPE_SSE}
\end{figure*}

The previous example clearly shows that the SOPE method can provide similar results to the Kalman filter, even when the hyperparameters have been selected on another example. However, the Kalman filter does not assume anything about the smoothness of the parameters being estimated. Therefore, it is natural to wonder how would the SOPE method behave in the presence of abrupt changes such as a discontinuity.

\subsection{Robustness of the SOPE approach}

In order to assess the robustness of the SOPE method to abrupt changes, tv-VAR model order misspecification and to model misspecification in general, we carry on extensive simulation studies. The following scenarios are proposed:
\begin{enumerate}
    \item The smoothness assumption is not always respected (e.g., presence of discontinuities)
    \item The time varying VAR model order $K$ is misspecified (over estimated and under estimated)
    \item The entire model is misspecified (e.g., mixture of AR(2) processes)
\end{enumerate}

\subsubsection{Abrupt parameter changes}

\begin{figure*}[p]
    \centering
    \includegraphics[width=.9\textwidth]{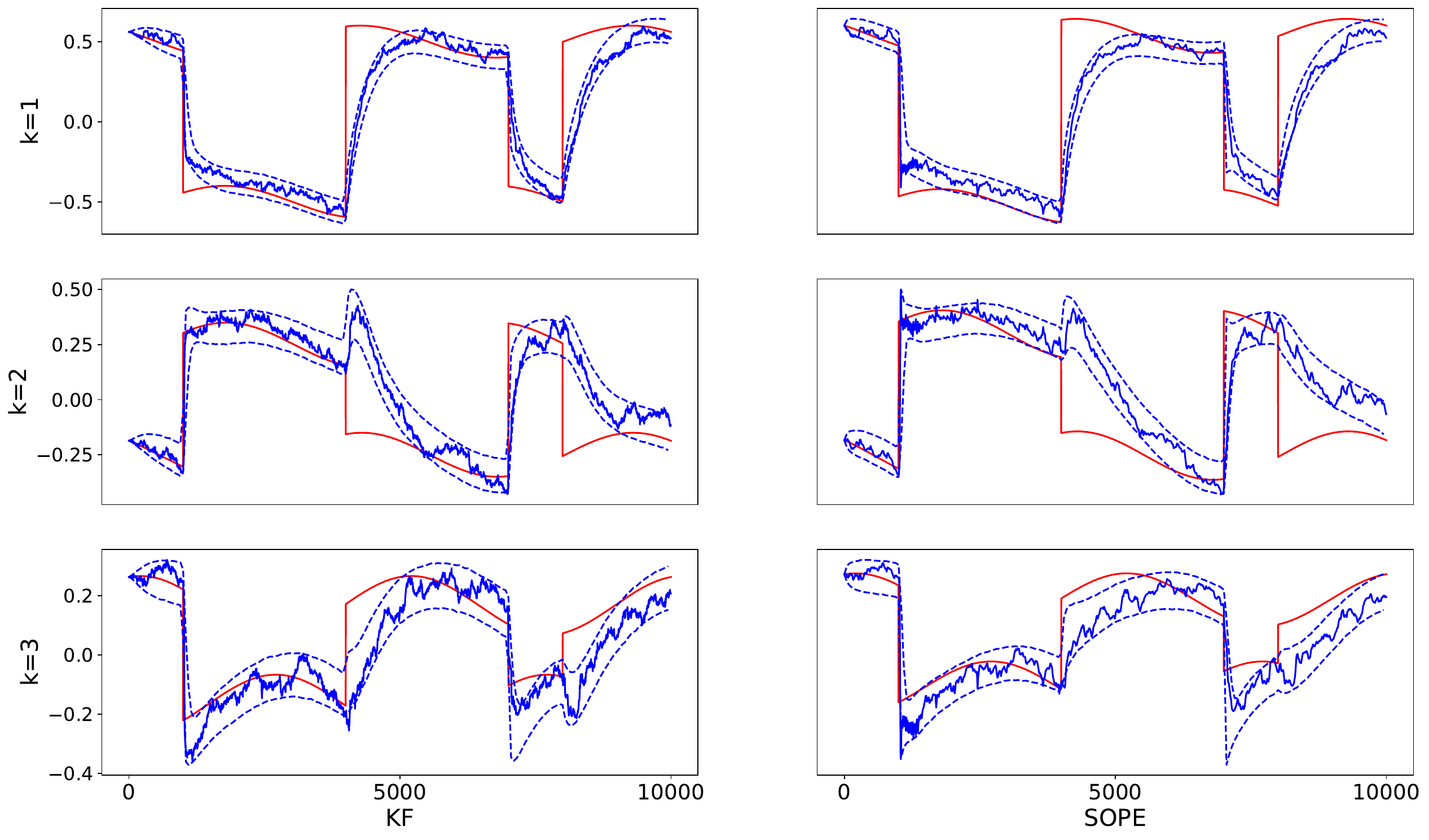}
    \caption{Parameter estimates for a tv-VAR model with with discontinuities, $P=3$ and $K=2$, pair of tetrodes $(2, 2)$. True parameters are in solid red lines, estimated parameters are in solid blue lines and $2,5$, $50$ and $97.5$ percentiles (based on $5000$ samples) are in doted blue lines. KF on the left column and SOPE on the right column.}
    \label{fig:SOPE_vs_KF_DISC}
\end{figure*}

In the following example, we propose to compare the KF and the SOPE over 5000 repetitions in the setting where tv-VAR model includes four discontinuities (at times $t=1000, t=4000, t=7000$ and $t=8000$) for some discontinuous parameters, which are $\Phi_{1}(t)_{2,2}, \Phi_{2}(t)_{2,2}, \Phi_{3}(t)_{2,2}$. Figure~\ref{fig:SOPE_vs_KF_DISC} shows how both approaches estimate the discontinuous parameter.

It is is clear from the figure above that the SOPE method is flexible enough to provide a wide range of estimates, from very smooth to very rough depending on the choice of the penalization hyperparameter. The SOPE has the additional advantage of being computationally fast which enables us to get similar quality results even in higher dimensions at a reasonable computational cost unlike the Kalman filter approach.

\subsubsection{Vector autoregression order misspecification}

In order for an estimation approach to be robust it needs to behave reasonably well when the order of the model is misspecified. In practice it is very difficult to approximate the order of a VAR model, and even more of a tv-VAR model.

\begin{figure*}[ht]
    \centering
    \includegraphics[width=.9\textwidth]{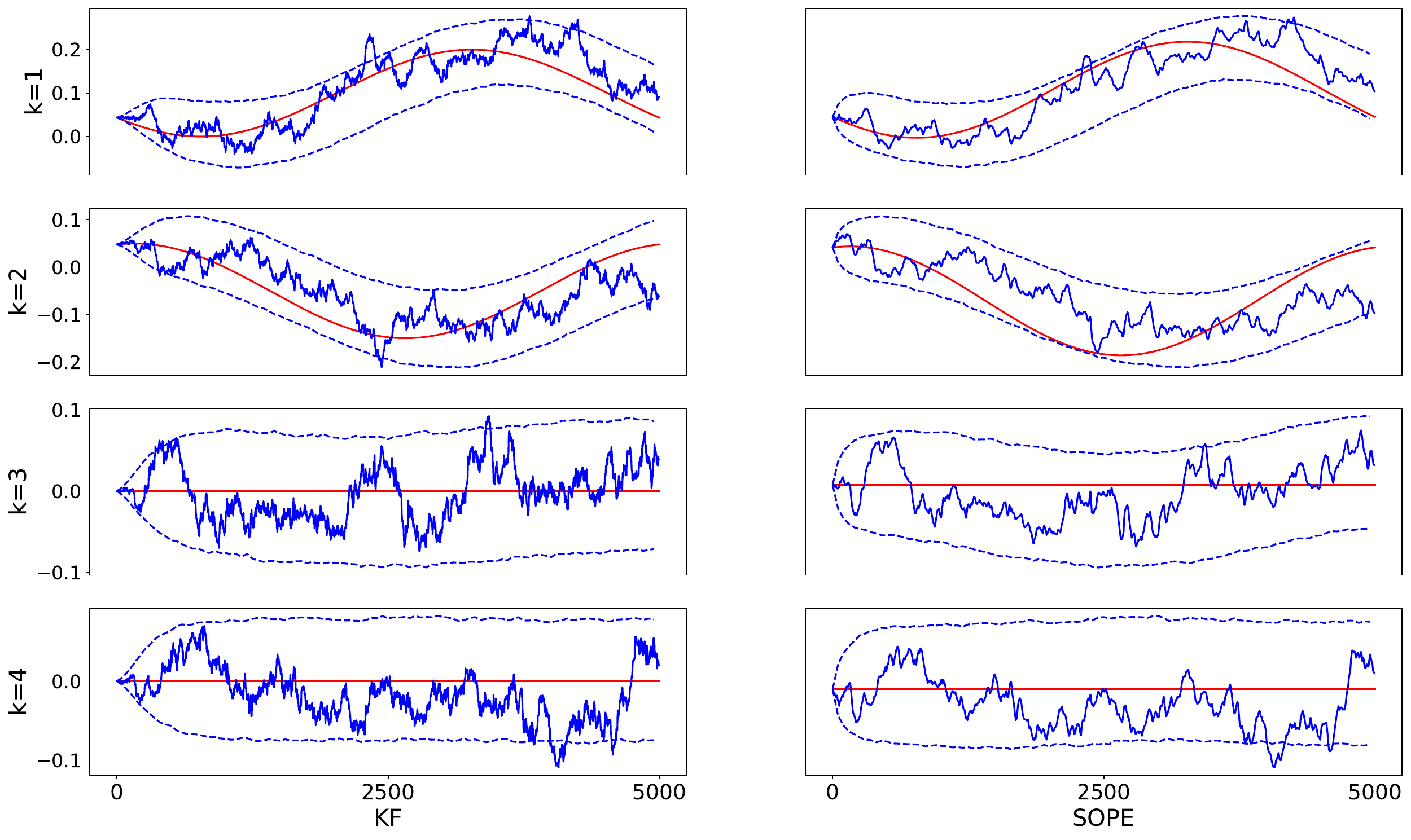}
    \caption{Parameter estimates for a tv-VAR model with with model misspecification (larger K), $P=4$, $K=2$ and $\widehat{K}=4$, pair of tetrodes $(1, 3)$. True parameters are in solid red lines, estimated parameters are in solid blue lines and $2,5$ and $97.5$ percentiles (based on $5000$ samples) are in doted blue lines. KF on the left column and SOPE on the right column.}
    \label{fig:SOPE_vs_KF__Model_Misspecification_larger}
\end{figure*}

In the following two examples we propose to fit a tv-VAR model ($P=4$) with the wrong order $K$ deliberately. First, when the order $K$ is over estimated, i.e., $\widehat{K}=4$ (fitted model in Equation \ref{Eq:fitted_VAR_model}) instead of $K=2$ (true model in Equation \ref{Eq:true_VAR_model}). Figure~\ref{fig:SOPE_vs_KF__Model_Misspecification_larger} shows the tv-VAR parameter estimates in addition to the confidence bands based on 5000 repetitions.
\begin{align}
    X(t) = \sum_{\ell =1}^K \Phi_{t,\ell}X(t-\ell) + E(t), \label{Eq:true_VAR_model}\\
    X(t) = \sum_{\ell =1}^{\widehat{K}} \Phi_{t,\ell}X(t-\ell) + E(t). \label{Eq:fitted_VAR_model}
\end{align}

\begin{figure*}[ht]
    \centering
    \includegraphics[width=.9\textwidth]{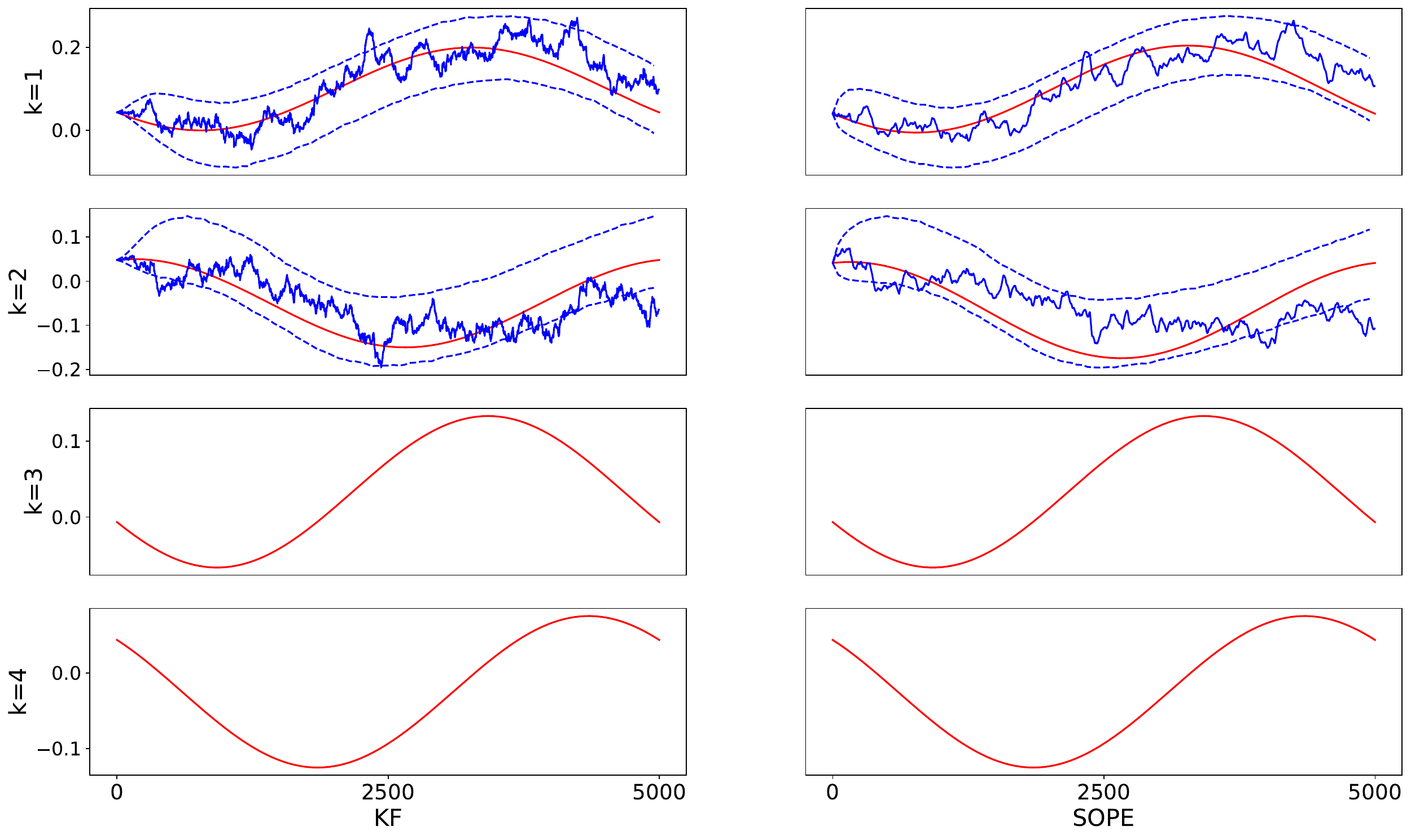}
    \caption{Parameter estimates for a tv-VAR model with with model misspecification (lower K), $P=4$, $K=4$ and $\widehat{K}=2$, pair of tetrodes $(1, 3)$. True parameters are in solid red lines, estimated parameters are in solid blue lines and $2,5$ and $97.5$ percentiles (based on $5000$ samples) are in doted blue lines. KF on the left column and SOPE on the right column.}
    \label{fig:SOPE_vs_KF__Model_Misspecification_lower}
\end{figure*}

Similarly, in the following example we propose to fit a tv-VAR model ($P=4$) with again the wrong order $K$. However, with the order $K$ being over estimated, i.e., $\widehat{K}=2$ instead of $K=4$. Figure~\ref{fig:SOPE_vs_KF__Model_Misspecification_lower} shows the estimates in addition to the confidence bands based on 5000 repetitions.

\begin{figure*}
    \centering
    \includegraphics[width=\linewidth]{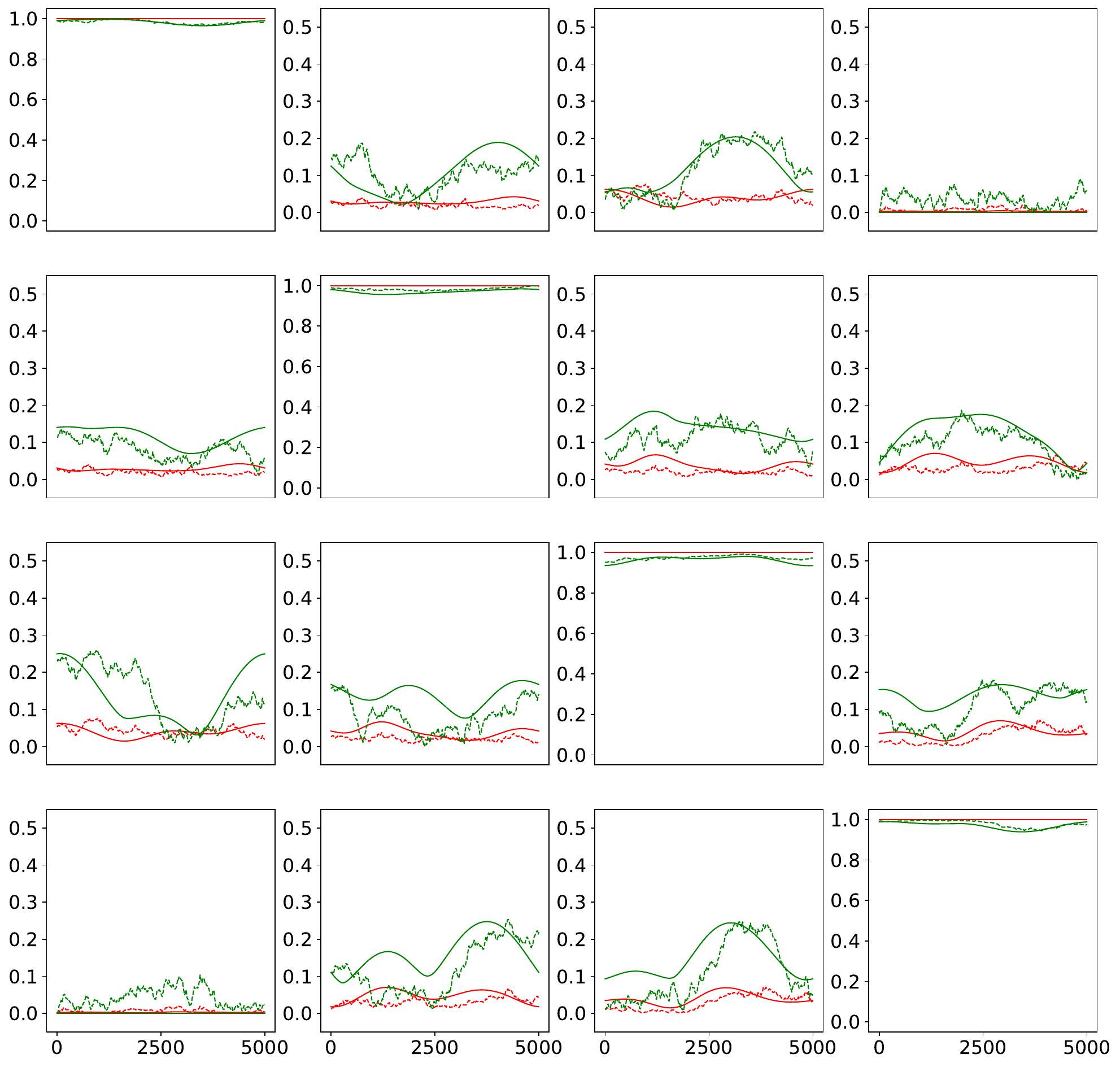}
    \caption{SOPE results for tv-VAR model under order misspecification $K=4$ instead of $K=2$: the plot in $i^{th}$ row and $j^{th}$ column represent the connectivity estimates between tetrode $i$ and $j$. Coherence is in red, while PDC is in green, solid lines represent the true quantity and dotted lines represent the estimates (for Delta band).}
    \label{fig:Conectivity_SIM_BETA_Model_Misspecification}
\end{figure*}

To see the impact of under estimating the tv-VAR model order on coherence and partial directed coherence, we fit the model then compute coherence and PDC, and report the results in Figure \ref{fig:Conectivity_SIM_BETA_Model_Misspecification}.

\subsubsection{Model misspecification}

The following example aims to present how the SOPE approach behaves when the entire model is misspecified. Often, brain signals can be modeled as a mixture of latent frequency-specific processes, as it has been proposed in \cite{BRAIN_SIGNALS_AR2S}. As such, it is reasonable to simulate data as a mixture of AR(2) processes. Thus, we propose the following formulation:

\begin{align}
    Y_1(t) = a(t) Z_L(t) + \big[ 1-a(t) \big] Z_H(t) + \epsilon_1(t) \\
    Y_2(t) = a(t) Z_L(t) + \big[ 1-a(t) \big] Z_H(t) + \epsilon_2(t) \\
    Y_3(t) = a(t) Z_H(t) + \big[ 1-a(t) \big] Z_L(t) + \epsilon_3(t) \\
    Y_4(t) = a(t) Z_H(t) + \big[ 1-a(t) \big] Z_L(t) + \epsilon_4(t)
\end{align}
Let $Z(t)$, $\epsilon(t)$ and $A(t)$ denote the following:
\begin{align}
    \epsilon(t) &= [\epsilon_1(t), \hdots, \epsilon_4(t)]' \\
    Z(t) &= [Z_L(t), Z_H(t)]' \\
    A(t) &= \begin{bmatrix}
                a(t) & 1-a(t) \\
                a(t) & 1-a(t) \\
                1-a(t) & a(t) \\
                1-a(t) & a(t)
            \end{bmatrix}
\end{align}
Then, the previous model can be written in the following compact form:
\begin{align}
    Y(t) &= A(t)Z(t) + \epsilon(t)
    \label{Eq:AR2_Mixture_Model}
\end{align}
where $A(t)$ is the time varying mixing matrix composed of the activation function $a(t) = \frac{1}{1 + \exp{[40(t-T/2)/T]}}$ and $Z(t)$ is the latent process composed of two AR(2) processes which spectrum is centered around low frequencies (100Hz) for $Z_L(t)$ and high frequencies (400Hz) for $Z_H(t)$, i.e., $[\phi_1^L=1.34, \phi_2^L=-0.69]$ and $[\phi_1^H=-1.34, \phi_2^H=-0.69]$. Thus, the first two components share a low frequency AR(2) process in the beginning of the experiment and a high frequency AR(2) process at the end of the experiment. However, the last two components share a high frequency AR(2) process in the beginning of the experiment and a low frequency AR(2) process at the end of the experiment, and . 

\begin{figure*}
    \centering
    \includegraphics[width=\linewidth]{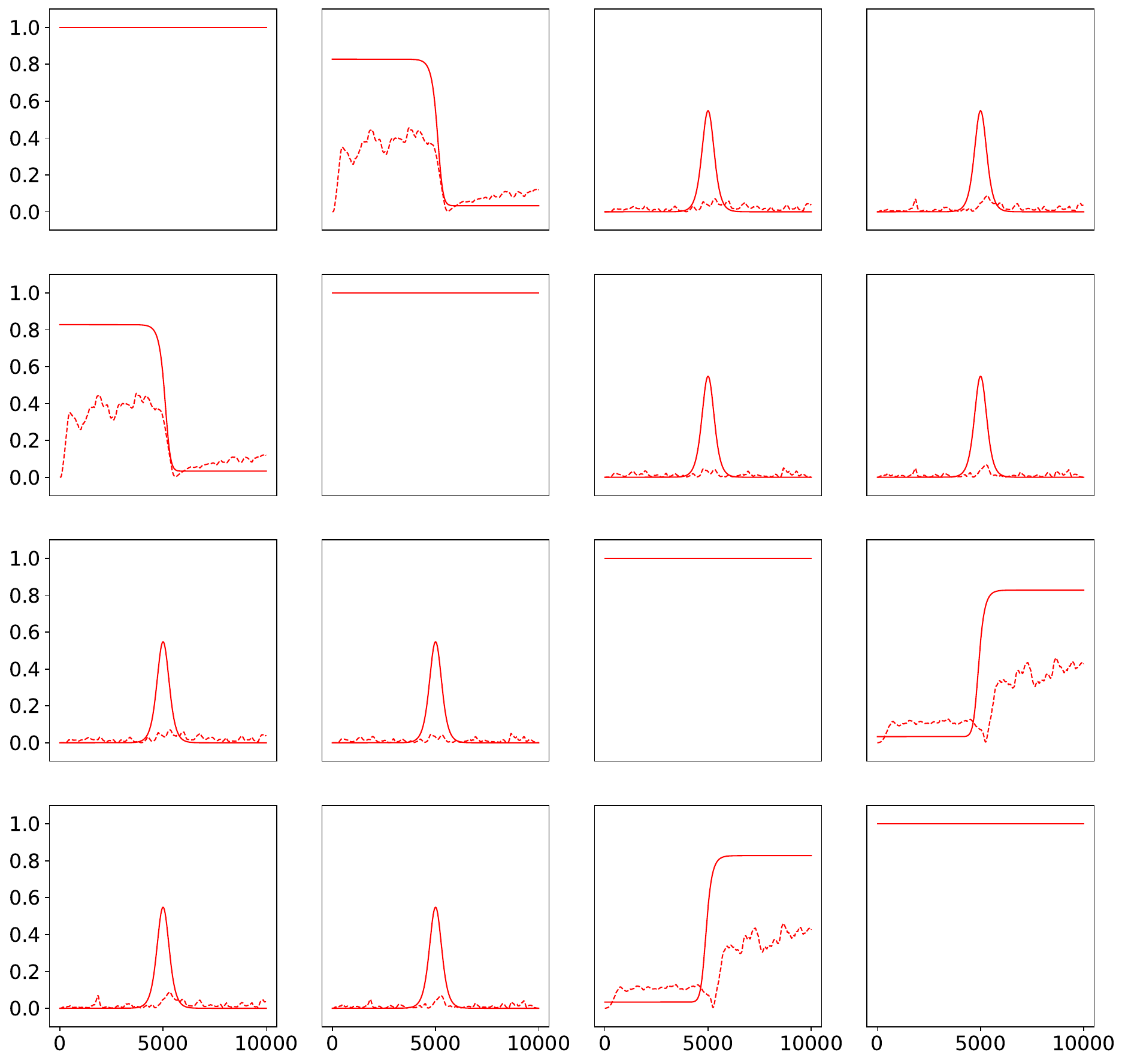}
    \caption{True and estimated time varying low frequency (50-150Hz) average coherence based on the tv-VAR model parameters with P=4, K=5, with initialization at zero. Solid lines represent the true coherence, while dotted lines represent the estimates.}
    \label{fig:Conectivity_Mixture_AR2_LF}
\end{figure*}

\begin{figure*}
    \centering
    \includegraphics[width=\linewidth]{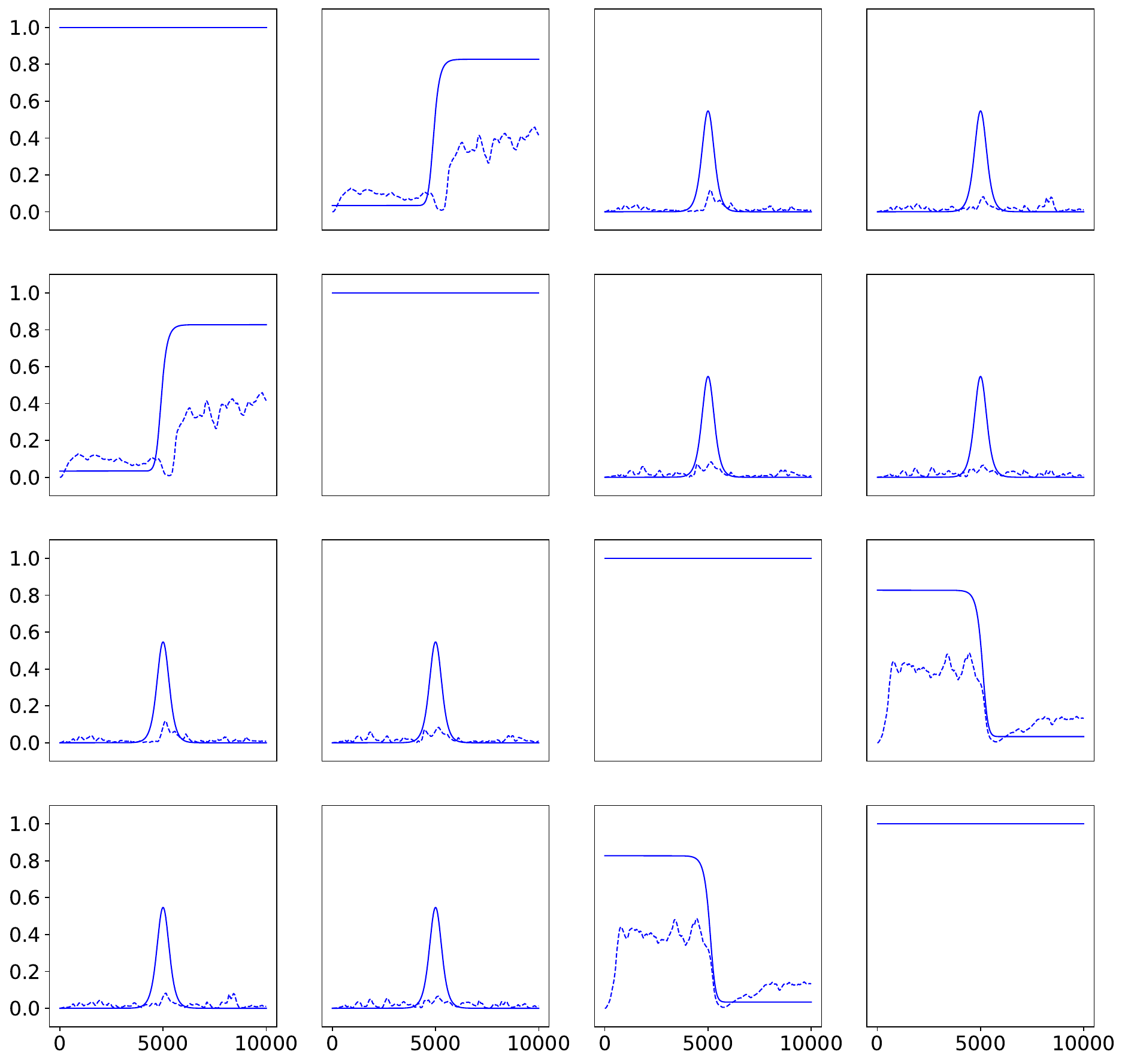}
    \caption{True and estimated time varying high frequency (350-450Hz) coherence based on the tv-VAR model with P=4, K=5, with initialization at zero. Solid lines represent the true coherence, while dotted lines represent the estimates.}
    \label{fig:Conectivity_Mixture_AR2_HF}
\end{figure*}

After generating the data from the previous model as defined in Equation \ref{Eq:AR2_Mixture_Model}, a tv-VAR model of dimension $P=4$ and order $K=5$ is fitted then the coherence is estimated/computed based on the estimated parameters and the true model. The results are reported in Figures \ref{fig:Conectivity_Mixture_AR2_LF} and \ref{fig:Conectivity_Mixture_AR2_HF} respectively for low and high frequency coherence.

Despite the fact that the coherence estimates in Figures \ref{fig:Conectivity_Mixture_AR2_LF} and \ref{fig:Conectivity_Mixture_AR2_HF} are far from the true coherence, the SOPE approach can capture the phase transition in the mixture of AR(2) processes quite clearly. Indeed, in low frequency coherence the estimates display a clear transition from a low plateau to a higher one for the first tow components, as can be seen in Figure \ref{fig:Conectivity_Mixture_AR2_LF}. However, for the last two components we observe the opposite transition from high frequency to low frequency, which is exactly what we expect. Similarly, we make the reverse observations for the high frequency in Figure \ref{fig:Conectivity_Mixture_AR2_HF}.

\section{Online estimation of connectivity of rat local field potentials}

\begin{figure*}[p]
\centering
    \begin{minipage}{.5\textwidth}
      \centering
      \includegraphics[width=.5\linewidth]{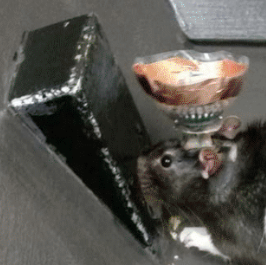}
      \caption{Rat with the implanted tetrodes.}
      \label{fig:rat_tetrodes}
    \end{minipage}%
    \begin{minipage}{.45\textwidth}
      \centering
      \includegraphics[width=.75\linewidth]{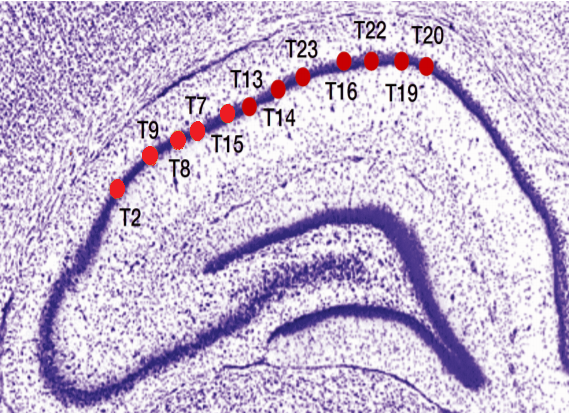}
      \caption{Tetrodes location in the hippocampus.}
      \label{fig:rat_tetrodes_location}
    \end{minipage}
    \label{fig:rat}
\end{figure*}

The ability to remember the order in which events occurred is fundamental to our daily life function. Considerable research shows that the hippocampus, a brain region strongly conserved across mammals, plays a crucial role in supporting that capacity. However, the underlying neural mechanisms are not well understood. To help address this issue, \cite{NON_SPATIAL_CODING} conducted an experiment in which they recorded neural activity from the CA1 subregion of the hippocampus as rats were tested on their memory of a sequence of five odors (lemon, rum, anise, vanilla, banana; see figure \ref{fig:rat_tetrodes} and \ref{fig:rat_tetrodes_location}).

In the context of closed-loop neurofeedback systems, it is critical to have access to real-time monitoring of the brain activity. Hence, fast processing of EEG signals is mandatory to control the appropriate feedback. Our goal in this application is to demonstrate that the SOPE method has the ability to capture the interesting dynamics of brain connectivity in real-time, which will allow others to examine the effects of adjusting experimental parameters, including stimulus intensity, to achieve the goals of the study or to provide the appropriate feedback for the system, see \cite{REAL_TIME_EEG_FC}.

In this analysis, 21 tetrodes were selected from 23 originally, namely, $T1, T2, \hdots, T23$ except $T11$ and $T17$. Since it is difficult to showcase the value of an online-method using static figures, a video was created to demonstrate this point. Real-time visualization of the estimated brain connectivity (for both coherence and PDC) network can be accessed in the following link: \href{https://www.dropbox.com/s/l0v5qmppgesjyay/LFP-Animation-00.mp4?dl=0}{real-time brain connectivity estimates}In the following, we present the analysis of the rat LFP recordings in order to illustrate the benefits gained from controlling the smoothness of the connectivity estimates, and how this approach can lead to the discovery of interesting patterns that could have been missed otherwise.

\begin{figure*}
    \centering
    \includegraphics[height=.25\linewidth, width=\linewidth]{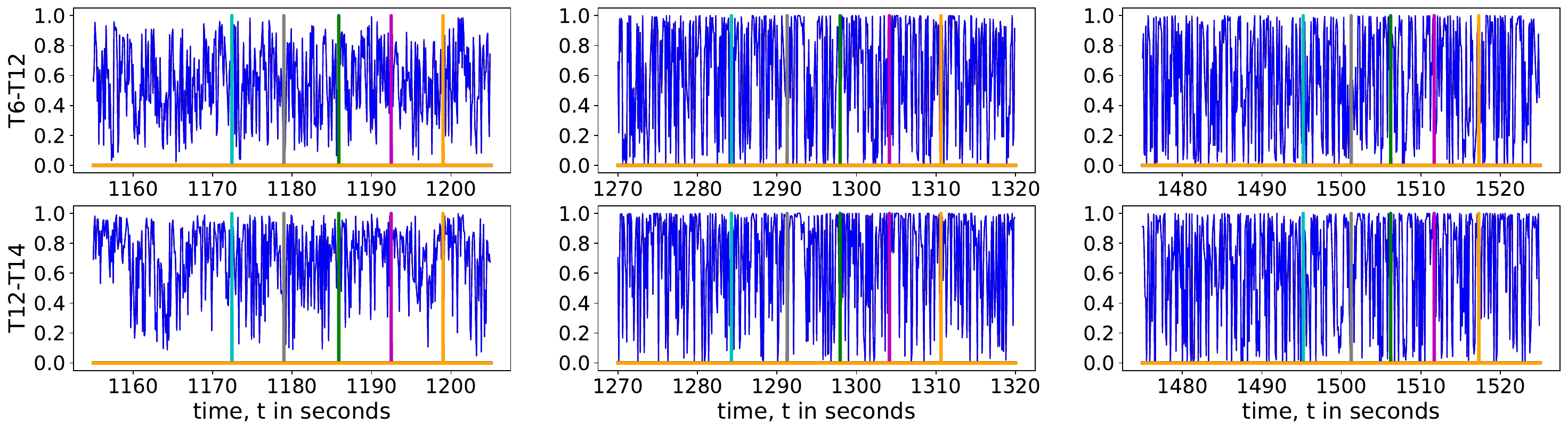}
    \caption{Coherence estimates using SOPE (slow gamma frequency band) during correct sequences of odors (lemon, rum, anise, vanilla, banana), vertical bars indicate the start of each odor. With regularization parameters $\alpha=500$ and $\beta=0.9$.}
    \label{fig:coherence_rough}
\end{figure*}

\begin{figure*}
    \centering
    \includegraphics[height=.25\linewidth, width=\linewidth]{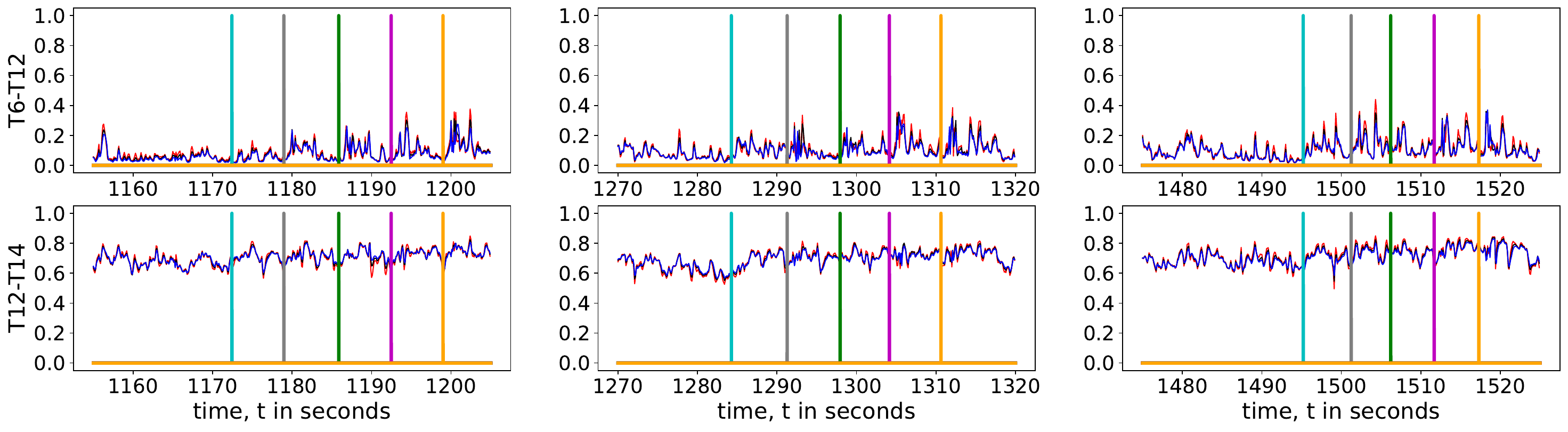}
    \caption{Coherence estimates using SOPE (slow gamma frequency band) during correct sequences of odors (in order: lemon, rum, anise, vanilla, banana), vertical bars indicate the start of each odor. With regularization parameters $\alpha=1500$ (red line)$, \alpha=20000$ (black line), $\alpha=25000$ (blue line) and $\beta=0.9$}
    \label{fig:coherence_smooth_slow_gamma}
\end{figure*}

\begin{figure*}
    \centering
    \includegraphics[height=.25\linewidth, width=\linewidth]{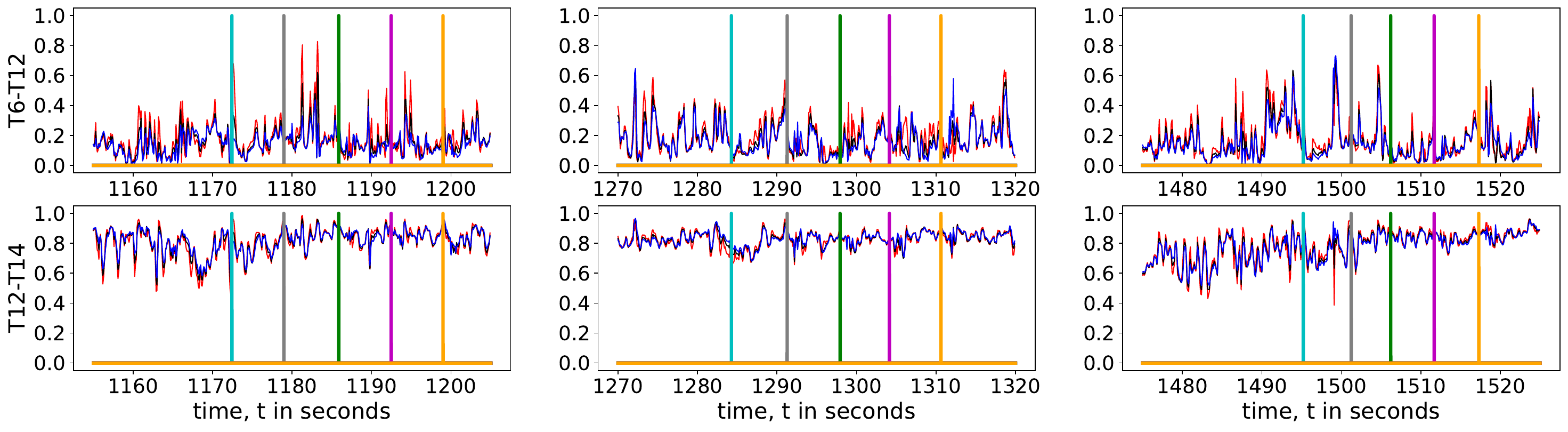}
    \caption{Coherence estimates using SOPE (theta frequency band) during correct sequences of odors (lemon, rum, anise, vanilla, banana), vertical bars indicate the start of each odor. With regularization parameters $\alpha=15000$ (red line)$, \alpha=2000$ (black line), $\alpha=25000$ (blue line) and $\beta=0.9$}
    \label{fig:coherence_smooth_theta}
\end{figure*}

As a starting point, a tv-VAR model of order $K=1$ is to fit to the LFP data where the dimension $P=21$, as it is the simplest model.
In the next step, we compute the connectivity measures presented in Section \ref{sbsec:connectivity_measures} (coherence and partial directed coherence) for the slow gamma (20-40 Hz) and theta-alpha (4-12 Hz) oscillation bands. As previously mentioned, these measures of connectivity are very sensitive to the variance of the observation noise. Therefore, a small penalization for the tv-VAR parameters estimator's roughness will lead to a poor connectivity estimator. Figure \ref{fig:coherence_rough} clearly shows the poor quality of such estimates using small penalization coefficient ($\alpha\sim 500$).
However, when more adequate (higher) levels of smoothing are selected, the connectivity estimates for coherence seem to be strongly modulated by the sequence of odors for the slow gamma band, as can be seen in Figure \ref{fig:coherence_smooth_slow_gamma}for the T6-T12 pair. Nevertheless, this amplitude modulation does not seem to be present for lower frequency bands such as the theta-alpha band, as can be seen in Figure \ref{fig:coherence_smooth_theta}. Furthermore, the above mentioned amplitude modulation is not present between all pairs of tetrodes (e.g., T12-T14 pair), which suggests some specific patterns in the activity at the electrode tips that is driven by the sequence of odors (instead of potential electrophysiological artifacts or noise). The presence of this amplitude modulation of dependence between some tetrodes in the slow gamma band and not in the theta band supports the results found in \cite{NON_SPATIAL_CODING}.

 In both Figures \ref{fig:coherence_smooth_slow_gamma} and \ref{fig:coherence_smooth_theta}, it can be seen that the choice of the penalization parameter $\alpha$ does not strongly influence the estimated connectivity as long as it is in the same order of magnitude, i.e., $\alpha \in O(10^4)$. This observation provides information regarding the low sensitivity of the SOPE method to the choice of the regularization parameter $\alpha$.

The offline analysis of the tetrodes connectivity network presents some interesting dynamics after an odor is introduced. To analyze this activity, the average connectivity measures were computed for the following temporal milestones: immediately before the odor presentation (-250 ms) and immediately after the presentation (+250 ms). These values were then compared with the connectivity's $0.75$-quantile threshold based on the empirical distribution which is described as follows.

The quantiles are computed based on the empirical distribution of the connectivity that is observed during the entire epoch which lasts for 50 seconds, at 1kHz sampling rate ($\widehat{\mathcal{C}}(t)\text{ and }\widehat{\mathcal{PDC}}(t), \forall t$). For every pair of tetrodes, the connectivity values are computed and then sorted from lowest to highest, the corresponding quantiles $(0.50, 0.75, 0.90)$ are computed, see Figure \ref{fig:coherence_smooth_slow_gamma}, top left subfigure as an example, where the $0.75$-quantile for the T5-T20 tetrode pair is represented in dotted black line.

\begin{figure*}
    \centering
    \includegraphics[width=.9\linewidth]{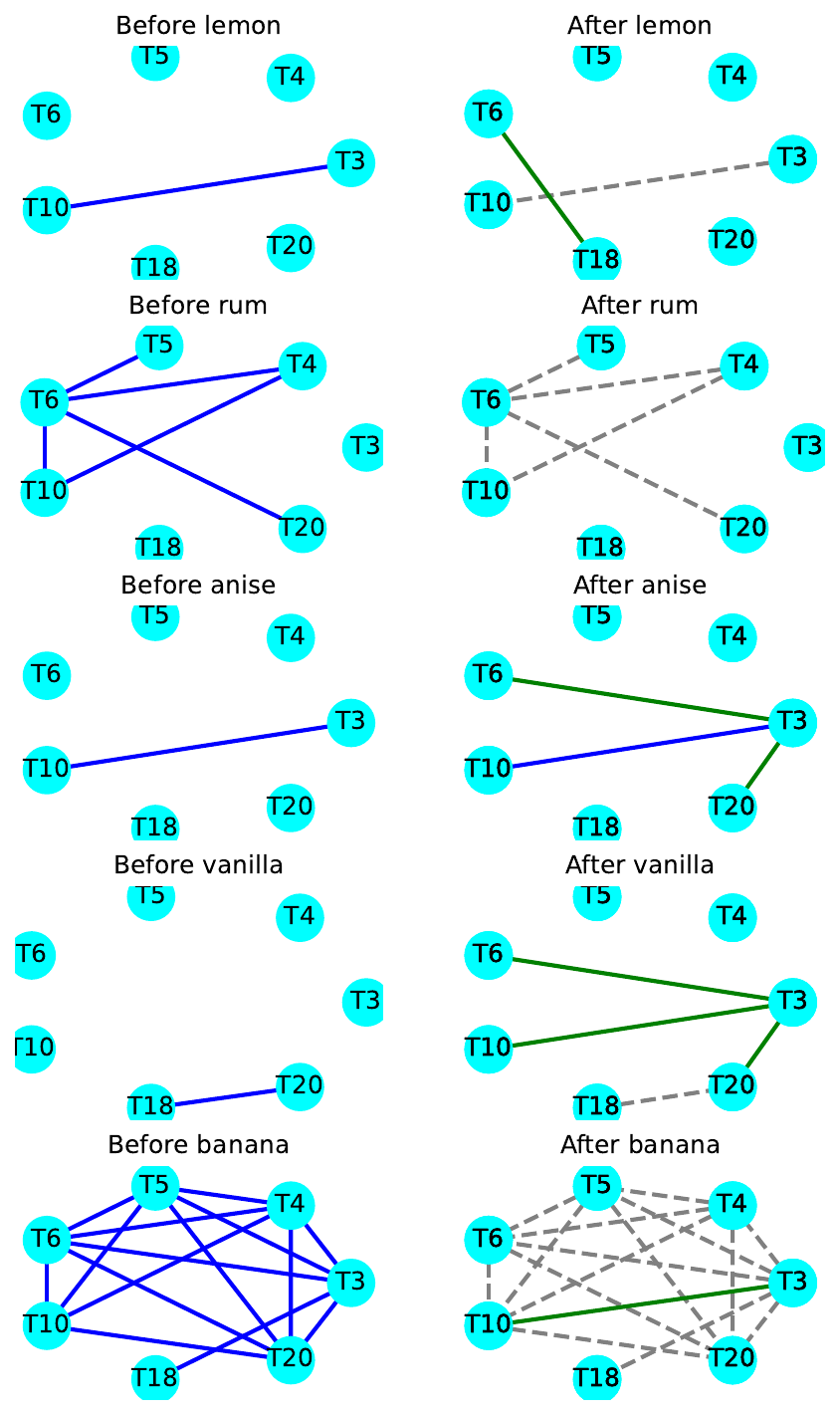}
    \caption{In the prior to odor presentation networks (based on the realizations at the start of the epoch) an edge is present if the high frequency coherence connectivity is higher than some threshold as defined by the 0.75 quantile (the quantile is computed based on the data before the odors are presented).
    On the post odor presentation networks (based on the realizations at the start of the epoch), a blue edge indicates that this connectivity was high prior to and it remained post odor presentation, a dotted gray edge indicates that this connectivity was high before and it is no longer the case, finally a green edge indicates that this connectivity was low before and it became high after the introduction of the banana odor.}
    \label{fig:network_coherence_high_freq_75}
\end{figure*}

\begin{figure*}
    \centering
    \includegraphics[width=.9\linewidth]{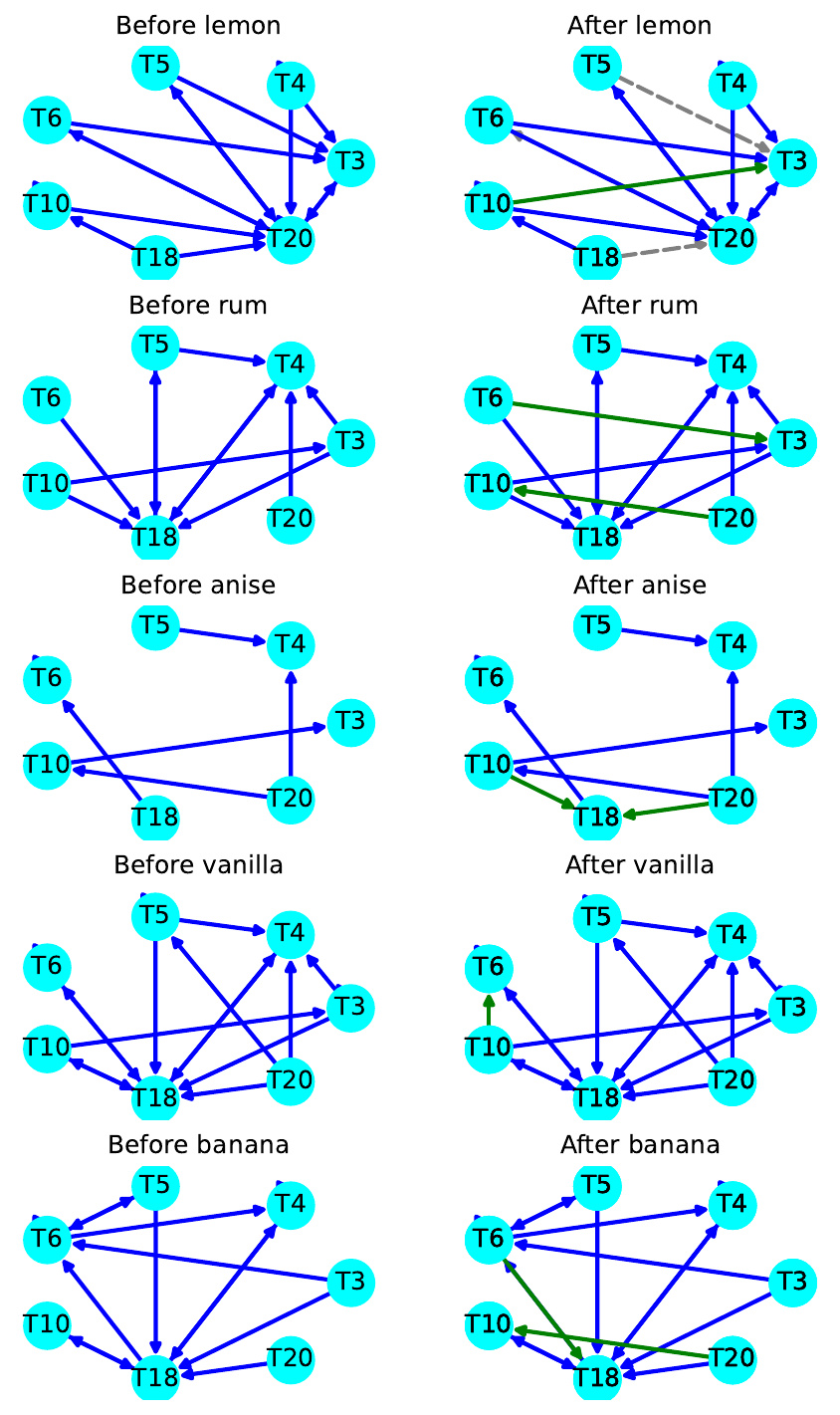}
    \caption{In the prior to odor presentation networks (based on the realizations at the start of the epoch) an edge is present if the high frequency PDC connectivity is higher than some threshold as defined by the 0.75 quantile (the quantile is computed based on the data before the odors are presented). On the post odor presentation networks (based on the realizations at the start of the epoch), a blue edge indicates that this connectivity was high prior to and it remained post odor presentation, a dotted gray edge indicates that this connectivity was high before and it is no longer the case, finally a green edge indicates that this connectivity was low before and it became high after the introduction of the banana odor.}
    \label{fig:network_pdc_high_freq_75}
\end{figure*}

\begin{figure*}
    \centering
    \includegraphics[width=.9\linewidth]{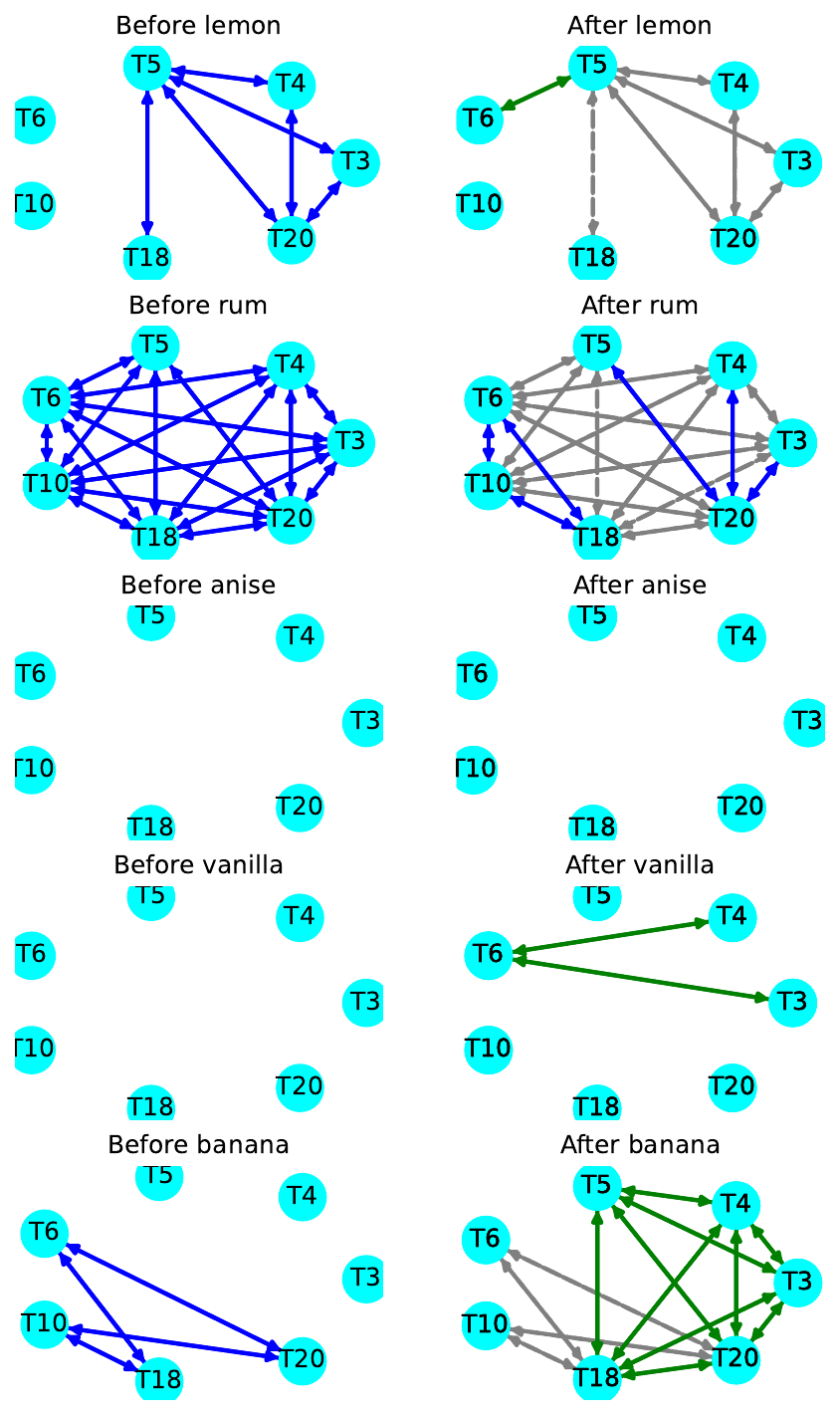}
    \caption{In the prior to odor presentation networks (based on the realizations at the start of the epoch) an edge is present if the low frequency coherence connectivity is higher than some threshold as defined by the 0.75 quantile (the quantile is computed based on the data before the odors are presented). On the post odor presentation networks (based on the realizations at the start of the epoch), a blue edge indicates that this connectivity was high prior to and it remained post odor presentation, a dotted gray edge indicates that this connectivity was high before and it is no longer the case, finally a green edge indicates that this connectivity was low before and it became high after the introduction of the banana odor.}
    \label{fig:network_coherence_low_freq_75}
\end{figure*}

\begin{figure*}
    \centering
    \includegraphics[width=.9\linewidth]{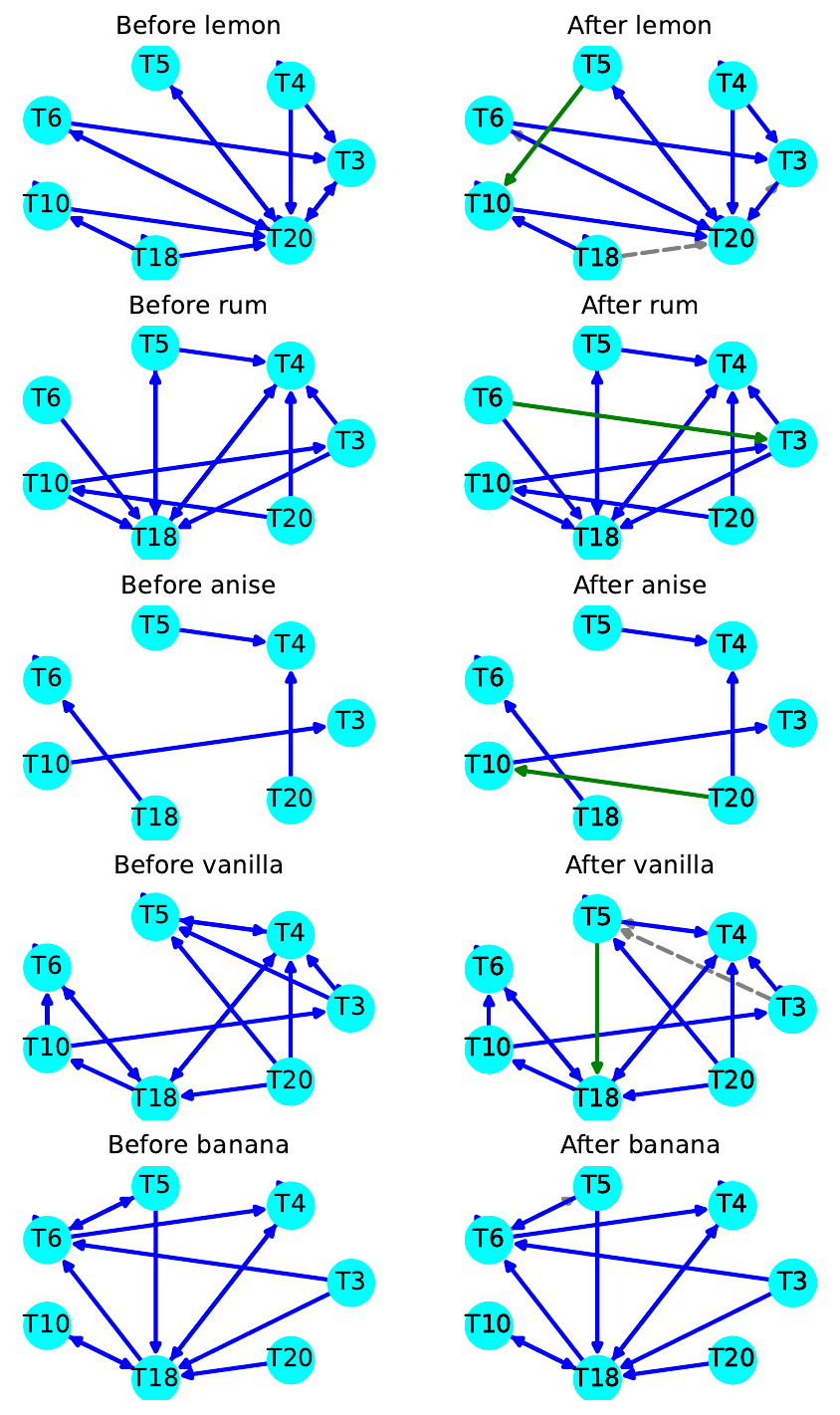}
    \caption{In the prior to odor presentation networks (based on the realizations at the start of the epoch) an edge is present if the low frequency PDC connectivity is higher than some threshold as defined by the 0.75 quantile (the quantile is computed based on the data before the odors are presented). On the post odor presentation networks (based on the realizations at the start of the epoch), a blue edge indicates that this connectivity was high prior to and it remained post odor presentation, a dotted gray edge indicates that this connectivity was high before and it is no longer the case, finally a green edge indicates that this connectivity was low before and it became high after the introduction of the banana odor.}
    \label{fig:network_pdc_low_freq_75}
\end{figure*}

We present bellow only the results for the $0.75$-quantile, the results for the $0.5$ and $0.9$ quantiles are kept for the Appendix, see Section \ref{subsec:quantile_networks}. In Figures \ref{fig:network_coherence_high_freq_75}, \ref{fig:network_pdc_high_freq_75}, \ref{fig:network_coherence_low_freq_75} and \ref{fig:network_pdc_low_freq_75}, the $T5$ and $T18$ tetrodes seem to play a central role as it appears to be associated with many changes in the graph structure. It also appears that the rum and banana odors seem to alter the connectivity between all tetrodes more often than the other odors. Results for other quantiles such as the $0.50$ and the $0.90$ are reported in the Appendix.

Coherence between the rat LFP tetrodes $i$ and $j$ can be viewed as a measure of the relative synchrony between the corresponding signals. More precisely, it is shown in \cite{OMBAO_BELLEGAN} to be the squared cross correlation between the components of the signals at some specific frequency band. On the other hand, partial directed coherence from tetrode $i$ to tetrode $j$ at some gven frequency band is a measure of the normalized (by the total amount of information flowing out from the $j$th tetrode) information flow from tetrode $j$ to tetrode $i$. Hence, those connectivity measures capture different facets of brain connectivity: the first one captures the relative synchrony between tetrodes and the second captures the portion of the directed flow of information from one tetrode to another. From Figures \ref{fig:network_coherence_high_freq_75}, \ref{fig:network_pdc_high_freq_75}, \ref{fig:network_coherence_low_freq_75} and \ref{fig:network_pdc_low_freq_75}, one can notice that coherence is a more conservative measure of connectivity (less connections in the networks), whereas partial directed coherence seems to be less conservative (more connections in the networks) by providing a more specific information about connectivity (directed and normalized).

The connectivity graphs in Figures \ref{fig:network_coherence_high_freq_75}, \ref{fig:network_pdc_high_freq_75}, \ref{fig:network_coherence_low_freq_75} and \ref{fig:network_pdc_low_freq_75}, were obtained respectively from high frequency coherence, high frequency partial directed coherence, low frequency coherence and low frequency partial directed coherence. For every odor (lemon, rum, anise, vanilla and banana) the average connectivity networks were computed 250ms before the presentation of the odor and 250ms after the presentation of the odor. To assess the significance of those measures, the comparison with quantile values of the empirical connectivity networks before the start of the epoch was made.
Therefore, an edge is drawn if the corresponding average connectivity was higher than the $75$-percentile threshold.
A green edge after the introduction of the odor, indicates an odor-induced increase in connectivity beyond the threshold. A gray edge after the introduction of the odor, indicates an odor-induced decrease in connectivity below the threshold.

The banana and rum odors introduced significant alterations in the connectivity structure, as can be seen in Figures \ref{fig:network_coherence_high_freq_75} and \ref{fig:network_pdc_low_freq_75}. The rum odor seem to inhibit both coherence and PDC, see Figures \ref{fig:network_coherence_high_freq_75} and \ref{fig:network_pdc_low_freq_75}. However, the banana seems to inhibit coherence at high frequency and increase PDC at low frequency, see \ref{fig:network_coherence_high_freq_75} and \ref{fig:network_pdc_low_freq_75}.

For the neuroscientist, it is important to be able to observe these alterations in brain connectivity as the odor stimuli are presented (rather than retrospectively) because this could allow for adaptive tuning of the odor intensity as needed. From the above analyses, it is clear that the SOPE method can successfully control the level of smoothness of the connectivity estimates. The SOPE provides access to smooth (robust to noise and artifacts) real-time estimates that can capture reliably the brain connectivity. Using the SOPE method will allow practitioners to have immediate feedback on the efficacy of the stimulus (type and intensity) that is being applied. This would allow the selection of the experiment settings before the completion of the data acquisition process. Real-time visualization of the rat's hippocampus connectivity network is shown 
in: \hyperlink{https://www.dropbox.com/s/hy35vspj57mso1m/LFP-Connectivity-Animation.mp4?dl=0}{real-time visualization of the rat's hippocampus connectivity network}.

\section{Conclusion}

We presented a new online parameter estimation method (SOPE) for locally stationary time series under the context of a tv-VAR model. It was shown that the proposed SOPE method can provide smooth and online estimates with a reasonable computational time that allows it to scale for higher dimensions unlike other competing methods. Furthermore,  the SOPE method is well motivated since it is based on a penalized least squares/likelihood approach. As demonstrated, the penalization term can have a Bayesian interpretation, with different penalization terms corresponding to different prior choices.

When compared to the Kalman filter, the SOPE method can provide very similar estimates, but with the significant advantage of being computationally more efficient, and thus it can be applied to higher dimensional time series. Furthermore, this new method provides a meaningful way to control the smoothness of the estimates using only two hyperparameters ($\alpha$ and $\beta$), which are much simpler to tune unlike the Kalman filter parameters. Asymptotically our approach has theoretical guaranties that indicate its asymptotic behaviour depending on the choice of the penalization coefficients. The interest behind this approach is to increase the robustness of the estimates and to stabilize the connectivity measures (which often involve non-smooth functions of the parameters e.g, Coherence, as it was shown in Figures \ref{fig:coherence_rough} and \ref{fig:coherence_smooth_theta}).
Using the SOPE method to estimate coherence and partial directed coherence between different tetrodes present in the CA1 region of the hippocampus, the SOPE method was able to detect the interesting dynamics in connectivity and the influence of the odor stimulus on the connectivity structure - in real-time. Thus, the SOPE will be a useful tool for experiments requiring the tracking of changes in connectivity in real-time for online manipulation of experimental parameters or circuit activity.

\newpage
\section{Appendix}

\subsection{Generalized SOPE} \label{sbsec:generalized_sope}

If the innovation components are not iid ($\Sigma_E \neq I$), the algorithm presented in section $2.3$ can be adapted similarly to the generalized least squares.
\begin{align}
    -log f\big(X(t)\big|b, U(t), \Sigma_E \big) \propto \big|\big| X(t) - b U(t) \big|\big|_{\Sigma_E^{-1}}^2
\end{align}
which in turn leads to the following problem:
\begin{align}
    &\widehat{\Phi}(t) = \underset{b\in \mathbb{R}^{P\times KP}}{\arg \max} f\big(X(t)\big|b, U(t) \big) f\big(b\big| \mathcal{I}_{t-1} \big) \\ \iff &\widehat{\Phi}(t) = \underset{b\in \mathbb{R}^{P\times KP}}{\arg \min} \big|\big| X(t) - bU(t) \big|\big|_{\Sigma_E^{-1}}^2 + P(b) \notag
\end{align}
Solving Equation $(26)$, is not straightforward. Assuming the covariance is known, a change of variable $$\Tilde{U} = \Big[\big(\Sigma_{E}^{-\frac{1}{2}} X(t-1)\big)', \hdots, \big(\Sigma_{E}^{-\frac{1}{2}} X(t-K)\big)'\Big]',$$ and $$\tilde{\Phi}(t) = \Sigma_E^{-\frac{1}{2}} \Phi(t) \big( I_K \otimes \Sigma_E^{\frac{1}{2}}  \big),$$, could be performed to decorrelate the time series components, which allows the problem to be solved in a similar way to the least squares problem, then another change of variable would be necessary to cancel the initial change of variable. In practice, the covariance matrix $\Sigma_E$ is unknown and consequently it has to be estimated simultaneously with the parameters of interest $\Phi(t)$ based on the residuals: 
\begin{align}
    \widehat{\Sigma}_E = \Sigma_T = \frac{T-1}{T} \Sigma_{T-1} + \frac{1}{T} R(T)R(T)' = \sum_{t=1}^T \frac{1}{T}R(t)R(t)'
\end{align}
this leads to the following general SOPE algorithm:
\vspace{5mm}

\newpage
\begin{algorithm}
  \caption{Smooth Online Parameter Estimation for tv-VAR models (general covariance)}\label{}
  \begin{algorithmic}[1]
    \Procedure{GetSmoothEstimates}{$X(1),\hdots, X(T)$}
    \State Initialize:
    \State $\widehat{\Phi}(K-1)$ = Least Squares
    \State $\widehat{\Phi}(K)$ = $\widehat{\Phi}(K-1)$
    \State $\Sigma_K = I$ (noise covariance)
    \State $\alpha \in (0, \infty)$, $\beta \in [0,\hdots, 1)$
    \For{$t=K+1,\hdots,T$}
    \State $\Tilde{X} = \Sigma_{t-1}^{-1/2} X(t)$
    \State $\Tilde{U} = \Big[\big(\Sigma_{t-1}^{-1/2} X(t-1)\big)', \hdots, \big(\Sigma_{t-1}^{-1/2} X(t-K)\big)'\Big]'$
    \State $\widehat{\tilde{\Phi}}(t) = \Big(\Tilde{X}\Tilde{U}'  + \lambda \Big[\widehat{\tilde{\Phi}}(t-1) + \beta \big(\widehat{\tilde{\Phi}}(t-1) - \widehat{\tilde{\Phi}}(t-2)\big)\Big]\Big)^{-1} \big(\Tilde{U}\Tilde{U}' + \lambda I\big)^{-1}$
    \State $\widehat{\Phi}(t) = \Sigma_{t-1}^{1/2} \widehat{\tilde{\Phi}}(t) \big( I_K \otimes \Sigma_{t-1}^{-1/2} \big)$
    \State $R(t) = X(t) - \widehat{\Phi}(t)U(t)$
    \State $\Sigma_{t} = \frac{t-1}{t}\Sigma_{t-1} + \frac{1}{t} R(t)R(t)'$
    \EndFor
    \EndProcedure
  \end{algorithmic}
\end{algorithm}

\subsection{Infill asymptotics} \label{sbsec:infill_asymptotics_appendix}

Let $h=\frac{1}{T} \underset{T\to \infty}{\to} dt$. In the following, we use the dot notation as a shortcut: $\dot{b}=\frac{d}{du}b$ and $\ddot{b}=\frac{d^2}{du^2}b$ etc., thus, for properly specified coefficients $h(T)$, the following limits are derived:

\vspace{3.2cm}
\begin{strip}
\begin{gather}
    \begin{cases}
       \lambda h^2 \underset{T \to \infty}{\to} c_1 \implies \lambda \big|\big| b(\frac{t}{T}) - b(\frac{t-1}{T}) \big|\big|_F^2 &= \lambda h^2 \big|\big| \frac{b(\frac{t}{T}) - b(\frac{t-1}{T})}{h} \big|\big|_F^2 \hspace{1.7cm} \to c_1 \big|\big| \dot{b}(u) \big|\big|_F^2 \\
       \lambda h^4 \underset{T \to \infty}{\to} c_2 \implies \lambda \big|\big| b(\frac{t}{T}) - 2 b(\frac{t-1}{T}) + b(\frac{t-2}{T}) \big|\big|_F^2 &= \lambda h^4 \big|\big| \frac{b(\frac{t}{T}) - 2 b(\frac{t-1}{T}) + b(\frac{t-2}{T})}{h^2} \big|\big|_F^2 \hspace{1.7cm} \hspace{-1.2cm} \to c_2 \big|\big| \ddot{b}(u) \big|\big|_F^2
    \end{cases}
\end{gather}
\end{strip}

\noindent hence, considering the overall estimation problem, one can deduce that the minimization problem of the sum of square errors will turn into a minimization of an integral provided that the penalization parameters grow at the adequate rate:

\begin{strip}
\begin{align}
    \sum_{t=1}^{T} \frac{1}{T} \Big[ \big|\big| X(t) - b(t)U(t) \big|\big|_2^2 + \lambda \big|\big| b(t) - b(t-1) \big|\big|_F^2 \Big] \to \int_{u=0}^{1} \big|\big| X(\lfloor uT \rfloor) - b(u)U(\lfloor uT \rfloor) \big|\big|_2^2 + c_1 \big|\big| \dot{b}(u) \big|\big|_F^2 du \\
    \sum_{t=1}^{T} \frac{1}{T} \Big[ \big|\big| X(t) - b(t)U(t) \big|\big|_2^2 + \lambda \big|\big| b(t) - 2b(t-1) + b(t-2) \big|\big|_F^2 \Big] \to \int_{u=0}^{1} \big|\big| X(\lfloor uT \rfloor) - b(u)U(\lfloor uT \rfloor) \big|\big|_2^2 + c_2 \big|\big| \ddot{b}(u) \big|\big|_F^2 du
\end{align}
\end{strip}

\noindent which leads to the following calculus of variations problems:

\begin{strip}
\begin{align}
    \widehat{\Phi} &= \underset{b\in \mathcal{C}^2([0,1],\mathbb{R}^{P\times KP})}{\arg \min} \int_{u=0}^{1} \big|\big| X(\lfloor uT \rfloor) - b(u)U(\lfloor uT \rfloor) \big|\big|_2^2 + c_1 \big|\big| \dot{b}(u) \big|\big|_F^2 du \\
    \widehat{\Phi} &= \underset{b\in \mathcal{C}^3([0,1],\mathbb{R}^{P\times KP})}{\arg \min} \int_{u=0}^{1} \big|\big| X(\lfloor uT \rfloor) - b(u)U(\lfloor uT \rfloor) \big|\big|_2^2 + c_2 \big|\big| \ddot{b}(u) \big|\big|_F^2 du
\end{align}
\end{strip}

\noindent in order to be rigorous here we need to state the boundary conditions of the problem. Of course, in practice we cannot have these boundary conditions since we need to solve this problem online. However, given previous estimates will automatically impose some boundary conditions.

The two problems in $(29)$ and $(30)$ involve a minimization over a space of functions. Now the problem can be stated in the framework of calculus of variations as follows. For penalty term in $(9)$ we asymptotically get:
\begin{align}
    \begin{cases}
        \mathcal{L}(u, b, \dot{b}) &= \big|\big| X(\lfloor uT \rfloor) - b(u)U(\lfloor uT \rfloor) \big|\big|_2^2+ \\ &\quad c_1 \big|\big| \dot{b}(u) \big|\big|_F^2 \\
        \mathcal{J}[b] &= \int_{0}^{1} \mathcal{L}(u, b, \dot{b}) du\\
        \widehat{\Phi} &= \underset{b\in B}{\arg \min } \hspace{.1cm} \mathcal{J}[b]
    \end{cases}
\end{align}
where the Lagrangian term is just the least squares term that corresponds to the likelihood of the observations  plus a penalty term for the roughness of the function, and $B$ is some smooth enough class of functions. To solve such problems, we consider a necessary condition for optimality also known as Euler-Lagrange equation of the involved Lagrangian:
\begin{align*}
    &\frac{\partial \mathcal{L}}{\partial b} - \frac{d}{du}\frac{\partial \mathcal{L}}{\partial \dot{b}} = 0 \Bigg|_{b=\widehat{\Phi}}  \\
    \iff &\ddot{\widehat{\Phi}}(u) = \frac{1}{c_1} \nabla_{\widehat{\Phi}} ||X(\lfloor uT \rfloor) - \widehat{\Phi}U(\lfloor uT \rfloor)||_2^2.
\end{align*}

\noindent similarly, using the second-order difference penalty term in $(10)$ will lead asymptotically to the formalized problem:
\begin{align}
    \begin{cases}
        \mathcal{L}(u, b, \dot{b}, \ddot{b}) &= \big|\big| X(\lfloor uT \rfloor) - b(u)U(\lfloor uT \rfloor) \big|\big|_2^2+ \\ &\quad c_2 \big|\big| \ddot{b}(u) \big|\big|_F^2 \\
        \mathcal{J}[b] &= \int_{0}^{1} \mathcal{L}(u, b, \dot{b}, \ddot{b}) du\\
        \widehat{\Phi} &= \underset{b\in B}{\arg \min} \hspace{.1cm} \mathcal{J}[b]
    \end{cases}
\end{align}
where again the Lagrangian term is just the least squares term that corresponds to the likelihood of the observations, plus a penalization term for the roughness (curvature) of the function. Similarly, the Euler-Lagrange equation becomes:

\begin{align*}
    &\frac{\partial \mathcal{L}}{\partial b} - \frac{d}{du}\frac{\partial \mathcal{L}}{\partial \dot{b}} + \frac{d}{du}\frac{\partial \mathcal{L}}{\partial \ddot{b}} = 0 \Bigg|_{b=\widehat{\Phi}} \\ \iff &\dddot{\widehat{\Phi}}(u) = -\frac{1}{c_2} \nabla_{\widehat{\Phi}} ||X(\lfloor uT \rfloor) - \widehat{\Phi}U(\lfloor uT \rfloor)||_2^2.
\end{align*}

To be able to use the results from the calculus of variations theory, some smoothness assumptions are necessary. in particular $b$ must be sufficiently smooth (differentiable) and the Lagrangian term must be a smooth function of $b$, $\dot{b}$ and of $\ddot{b}$, which is obviously the case here since we deal with a quadratic function of $b$, $\dot{b}$ and of $\ddot{b}$. The smoothness of $b$ can be controlled in simulation ($\mathcal{C}^2$, $\mathcal{C}^3$ etc.). In practice, however, this might be violated but we assume it is the case since we are concerned with locally stationary processes. The above conditions are necessary for deriving the Euler-Lagrange equation.

\hspace*{0pt}\vfill\hspace*{0pt}

\subsection{Parameter and connectivity matrices}
\label{subsec:full_param_connectivity_matrices}

\begin{figure}[H]
    \centering
    \includegraphics[width=.9\linewidth]{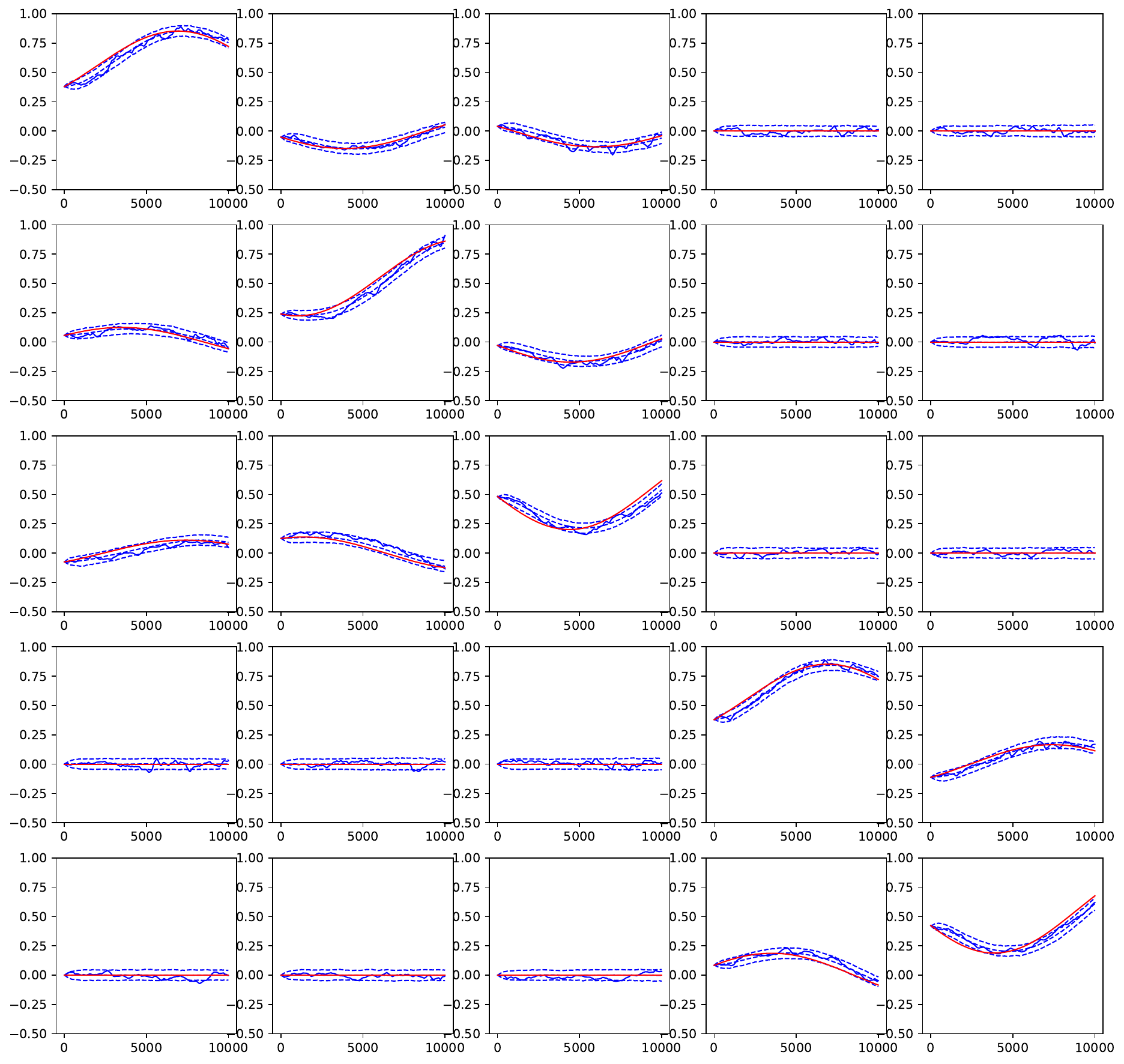}
    \caption{SOPE results for tv-VAR model: the plot in $i^{th}$ row and $j^{th}$ column represent the results for $\big(\Phi_1(t)\big)_{i, j}$ (red line) and $\big(\widehat{\Phi}_1(t)\big)_{i, j}$ (blue line), the 2.5, 50 and 97.5 percentiles (based on $B=1000$ samples) are represented in dotted blue lines.}
    \label{fig:PARAMS_SIM}
\end{figure}

\begin{figure}[H]
    \centering
    \includegraphics[height=.9\linewidth, width=.9\linewidth]{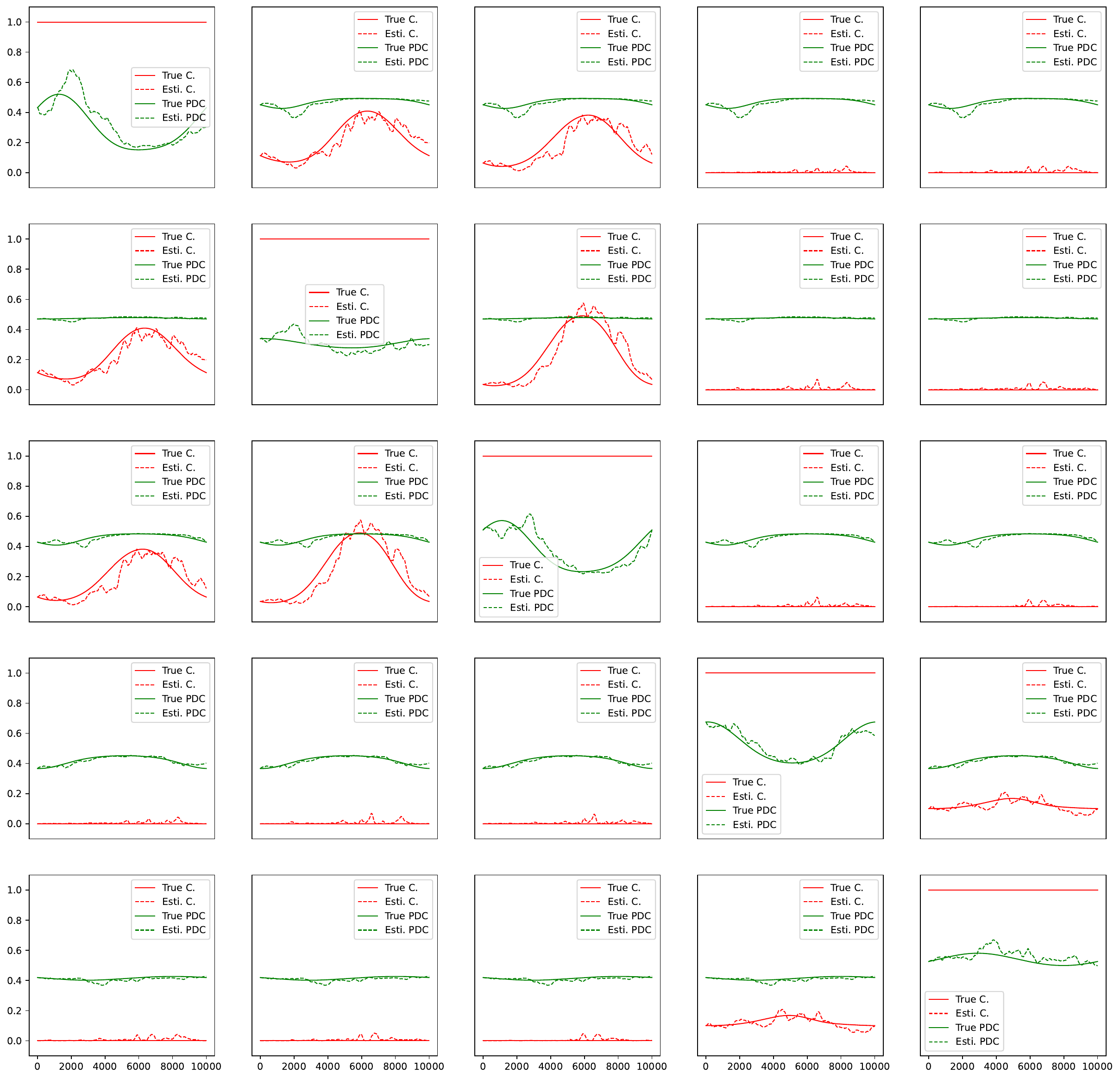}
    \caption{SOPE results for tv-VAR model: the plot in $i^{th}$ row and $j^{th}$ column represent the connectivity estimates between tetrode $i$ and $j$. Solid lines represent the true quantity and dotted lines represent the estimates (for Delta band).}
    \label{fig:Conectivity_SIM_DELTA}
\end{figure}

\subsection{Connectivity networks prior to and post odor presentation}
\label{subsec:quantile_networks}

Connectivity networks for $0.5$ and $0.95$ quantiles:

\begin{figure}[H]
    \centering
    \includegraphics[width=.99\linewidth]{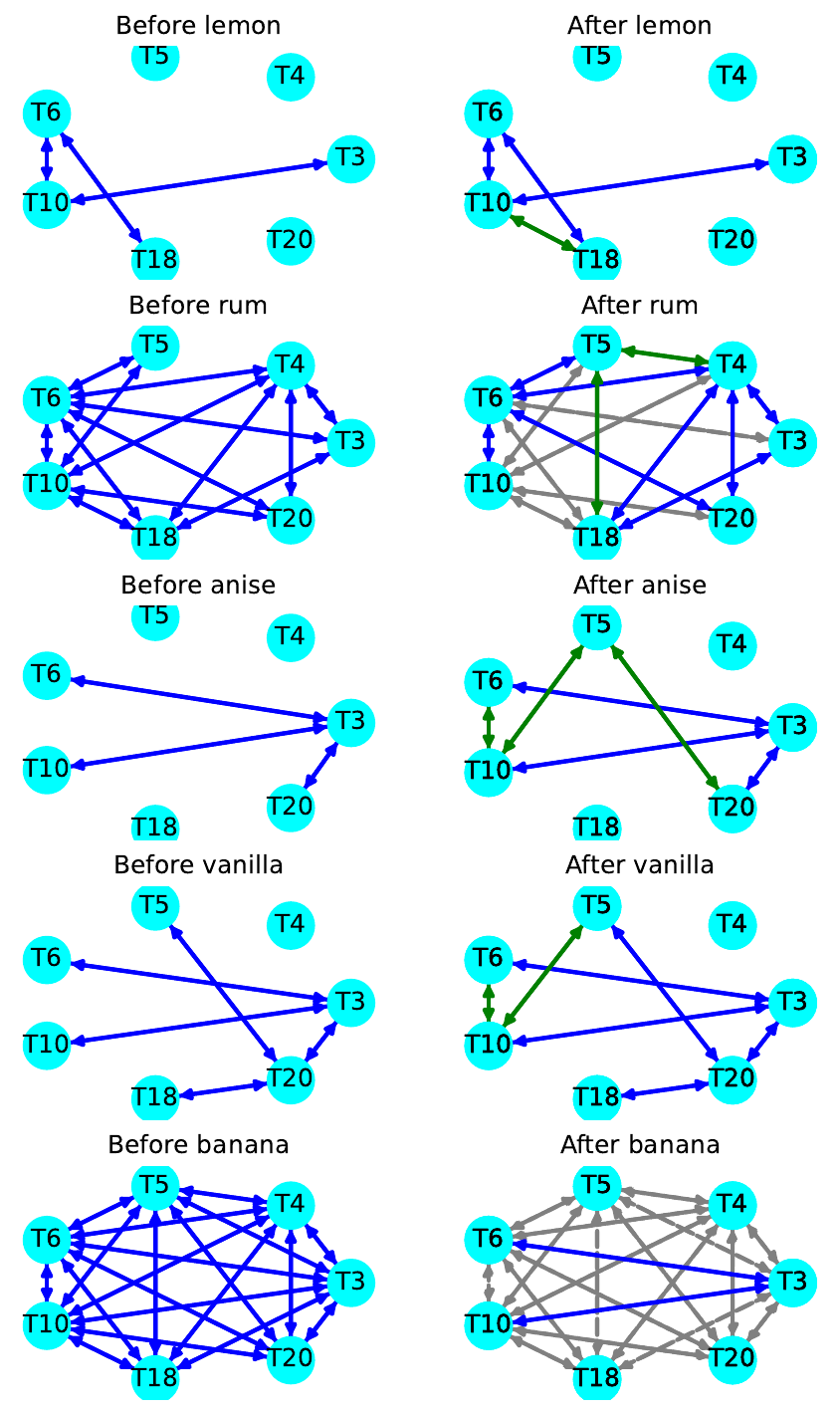}
    \caption{In the before-odor networks an edge is present if the high frequency coherence connectivity is higher than some threshold as defined by the 0.5 quantile. On the post odor presentation networks (based on the realizations at the start of the epoch), a blue edge indicates that this connectivity was high prior to and it remained post odor presentation, a dotted line gray edge indicates that this connectivity was high before and it is no longer the case and finally a green edge indicates that this connectivity was low before and it became high after the introduction of the smell.}
    \label{fig:network_coherence_high_freq_50}
\end{figure}
\hspace*{0pt}\vfill\hspace*{0pt}
\begin{figure}[H]
    \centering
    \includegraphics[width=.99\linewidth]{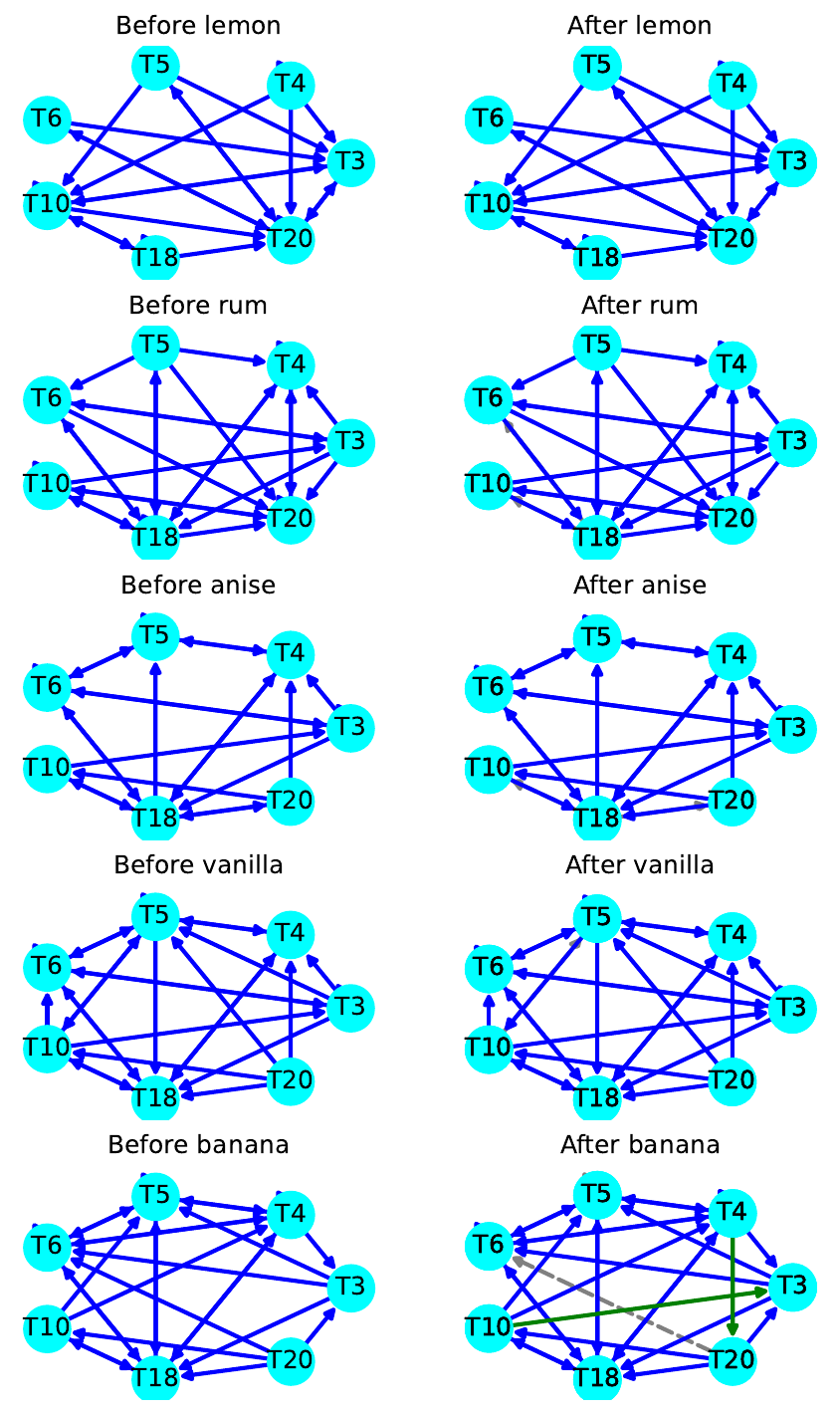}
    \caption{In the before-odor networks an edge is present if the high frequency PDC connectivity is higher than some threshold as defined by the 0.5 quantile. On the post odor presentation networks (based on the realizations at the start of the epoch), a blue edge indicates that this connectivity was high prior to and it remained post odor presentation, a dotted line gray edge indicates that this connectivity was high before and it is no longer the case and finally a green edge indicates that this connectivity was low before and it became high after the introduction of the smell.}
    \label{fig:network_pdc_high_freq_50}
\end{figure}
\begin{figure}[H]
    \centering
    \includegraphics[width=.99\linewidth]{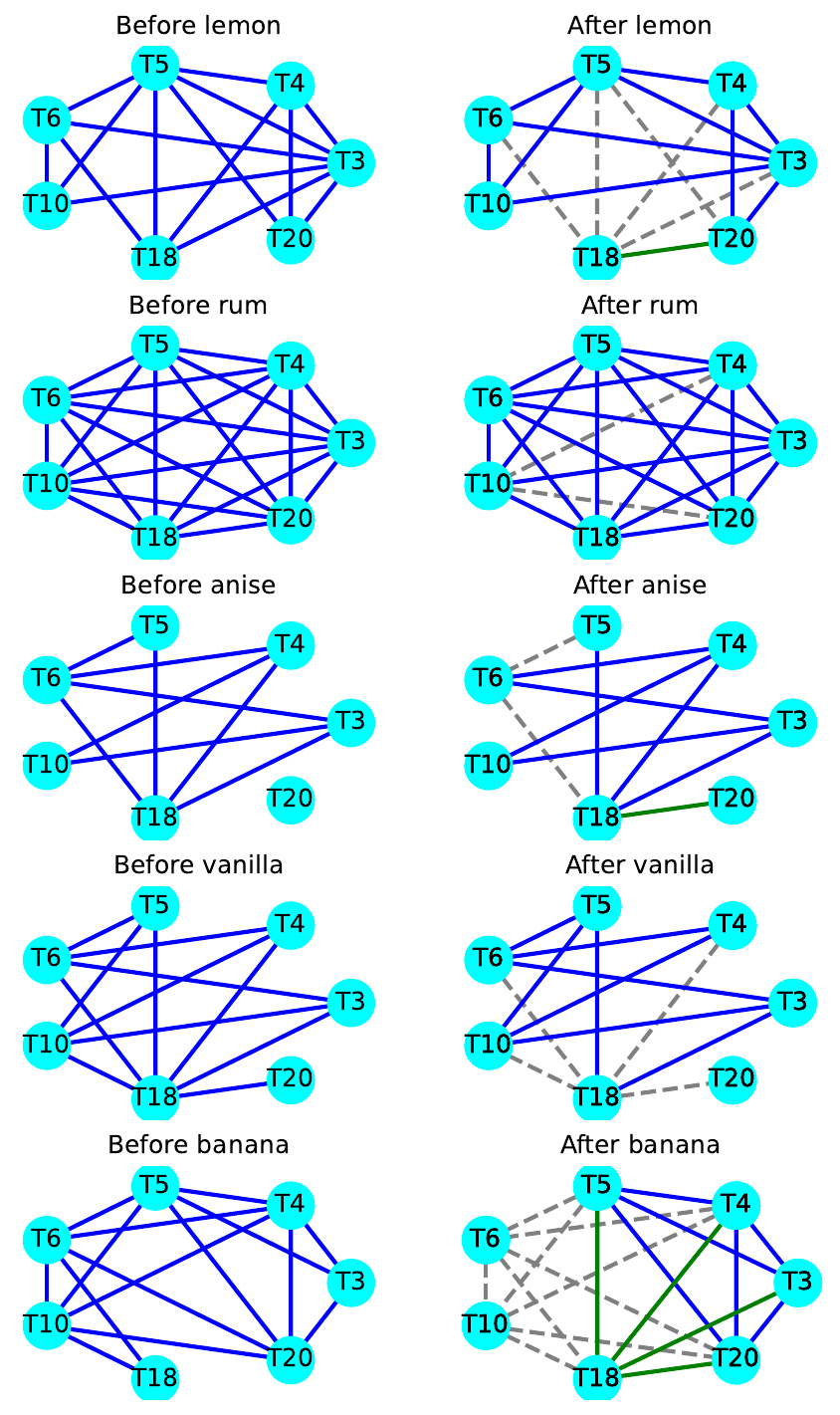}
    \caption{In the before-odor networks an edge is present if the high frequency coherence connectivity is higher than some threshold as defined by the 0.5 quantile. On the post odor presentation networks (based on the realizations at the start of the epoch), a blue edge indicates that this connectivity was high prior to and it remained post odor presentation, a dotted line gray edge indicates that this connectivity was high before and it is no longer the case and finally a green edge indicates that this connectivity was low before and it became high after the introduction of the smell.}
    \label{fig:network_coherence_low_freq_50}
\end{figure}

\begin{figure}[H]
    \centering
    \includegraphics[width=.99\linewidth]{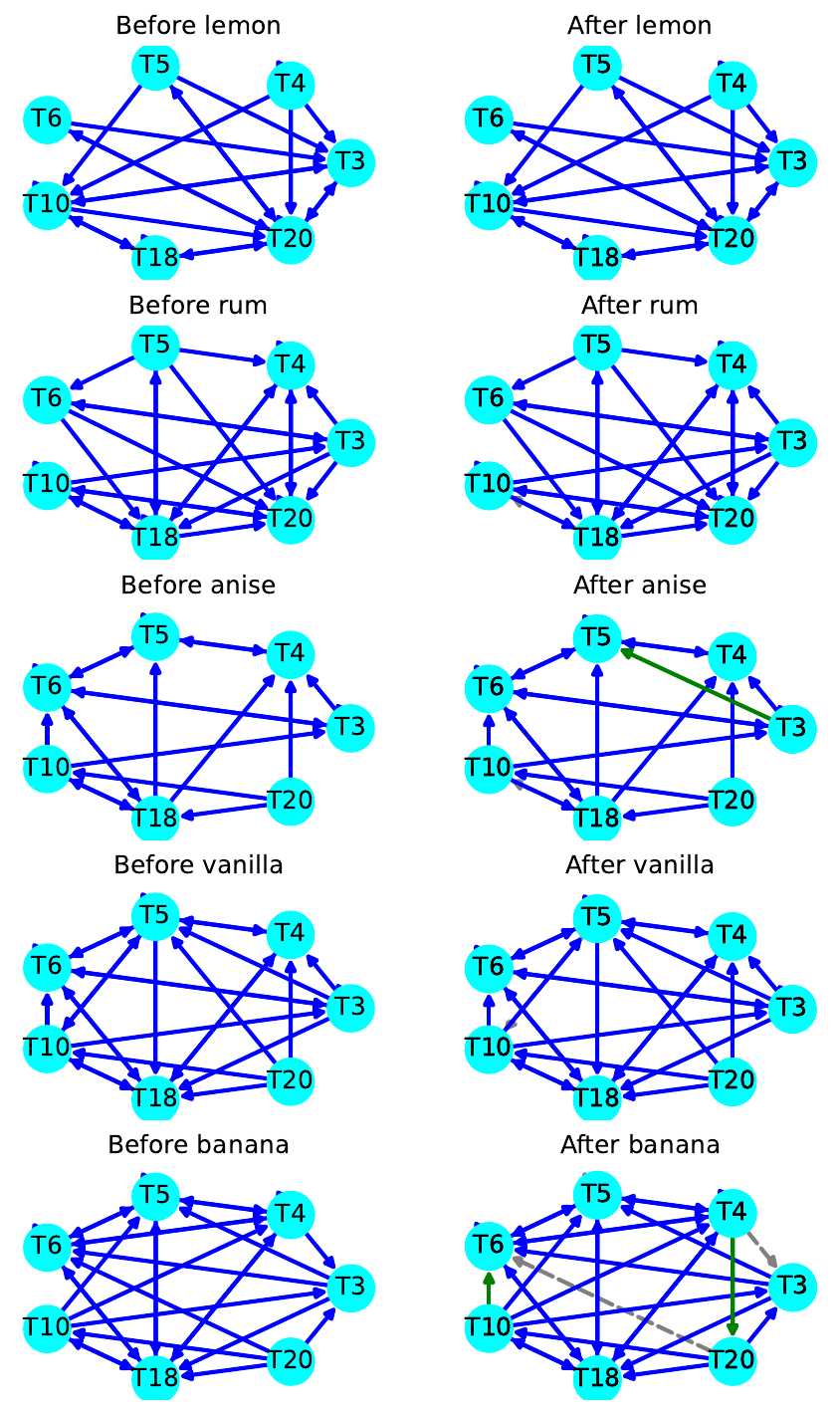}
    \caption{In the before-odor networks an edge is present if the low frequency PDC connectivity is higher than some threshold as defined by the 0.5 quantile. On the post odor presentation networks (based on the realizations at the start of the epoch), a blue edge indicates that this connectivity was high prior to and it remained post odor presentation, a dotted line gray edge indicates that this connectivity was high before and it is no longer the case and finally a green edge indicates that this connectivity was low before and it became high after the introduction of the smell.}
    \label{fig:network_pdc_low_freq_50}
\end{figure}

\begin{figure}[H]
    \centering
    \includegraphics[width=.99\linewidth]{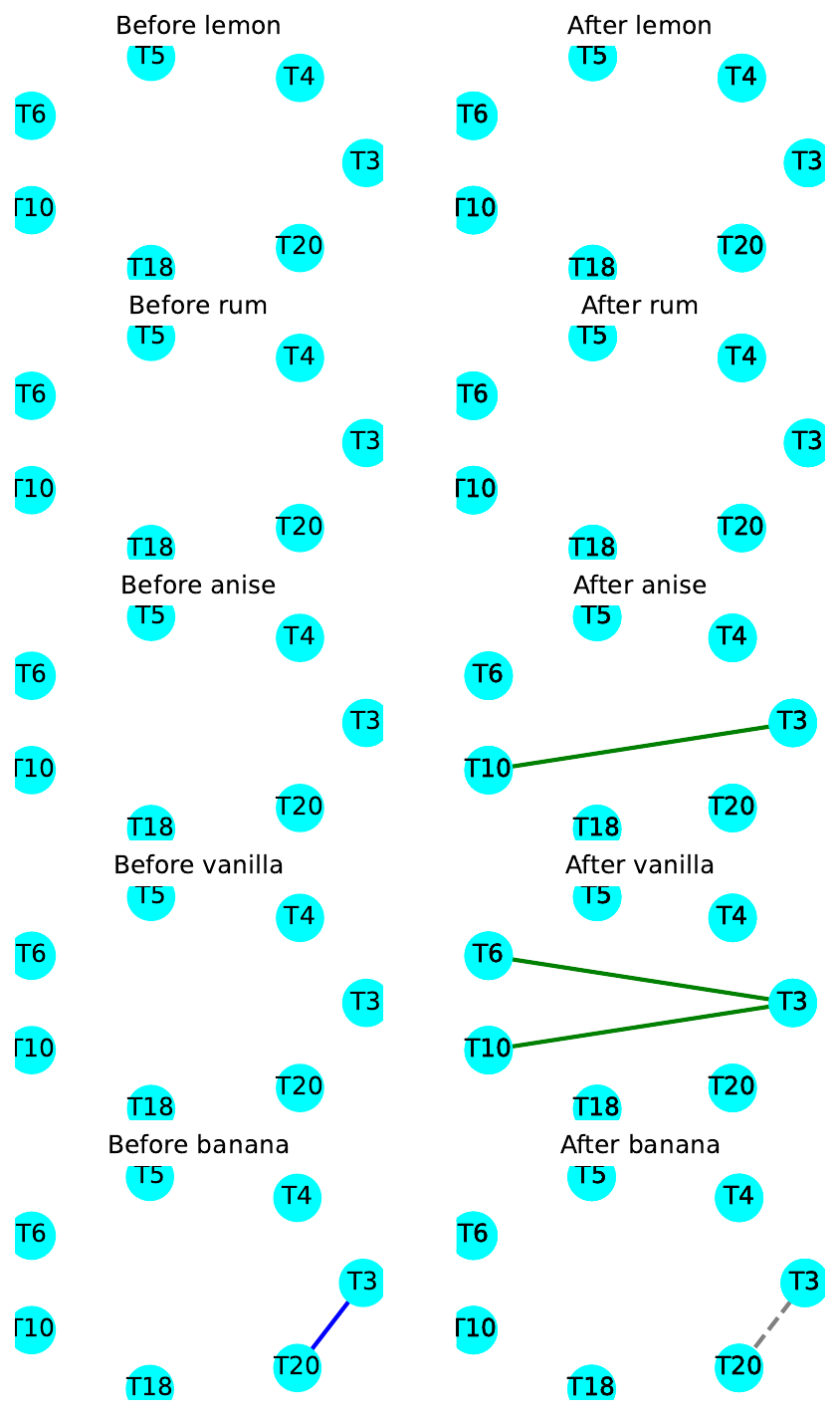}
    \caption{In the prior to odor presentation networks (based on the realizations at the start of the epoch) an edge is present if the high frequency coherence connectivity is higher than some threshold as defined by the 0.9 quantile. On the post odor presentation networks (based on the realizations at the start of the epoch), a blue edge indicates that this connectivity was high prior to and it remained post odor presentation, a dotted line gray edge indicates that this connectivity was high before and it is no longer the case and finally a green edge indicates that this connectivity was low before and it became high after the introduction of the smell.}
    \label{fig:network_coherence_high_freq_90}
\end{figure}

\begin{figure}[H]
    \centering
    \includegraphics[width=.99\linewidth]{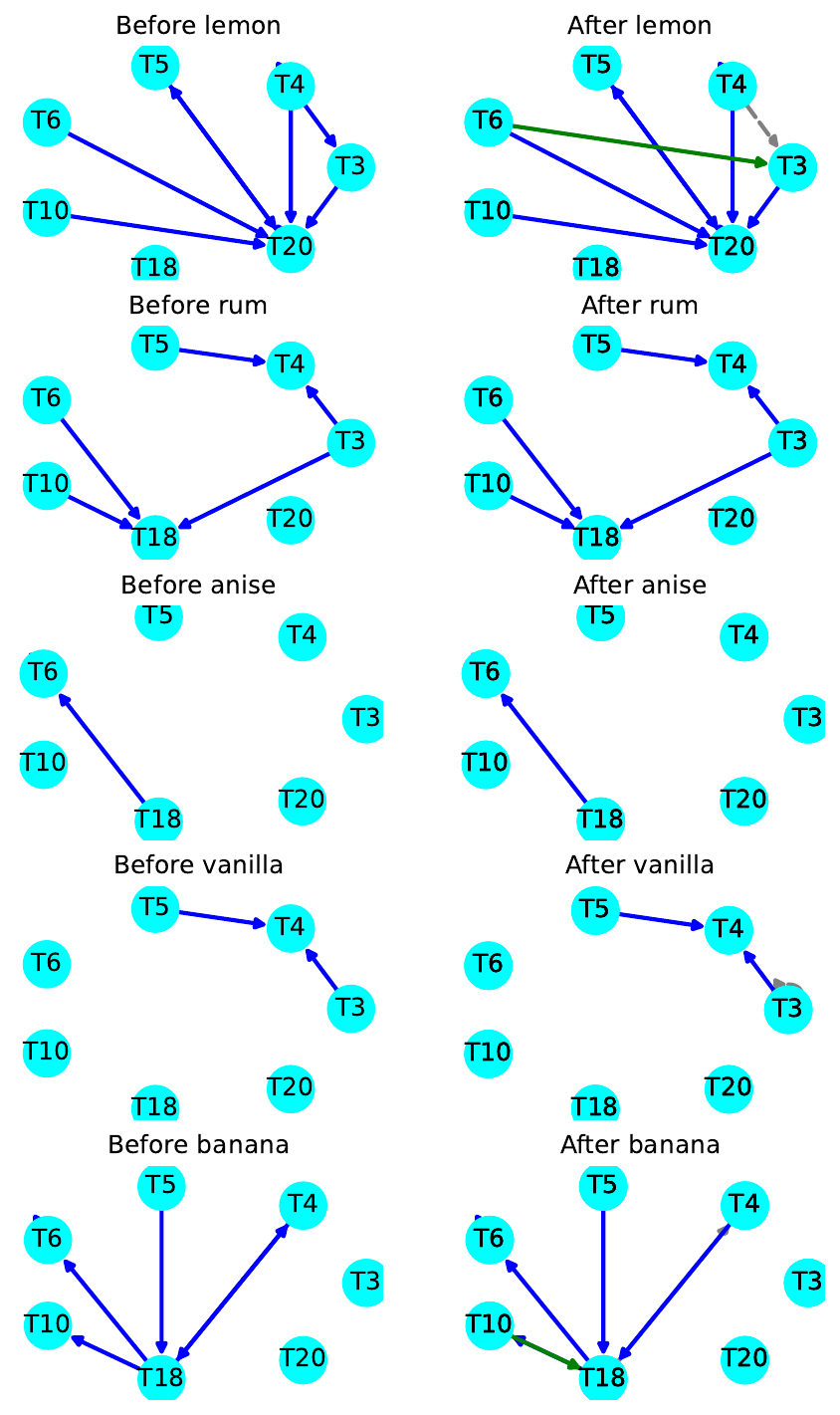}
    \caption{In the prior to odor presentation networks (based on the realizations at the start of the epoch) an edge is present if the high frequency PDC connectivity is higher than some threshold as defined by the 0.9 quantile. On the post odor presentation networks (based on the realizations at the start of the epoch), a blue edge indicates that this connectivity was high prior to and it remained post odor presentation, a dotted line gray edge indicates that this connectivity was high before and it is no longer the case and finally a green edge indicates that this connectivity was low before and it became high after the introduction of the smell.}
    \label{fig:network_pdc_high_freq}
\end{figure}
\begin{figure}[H]
    \centering
    \includegraphics[width=.99\linewidth]{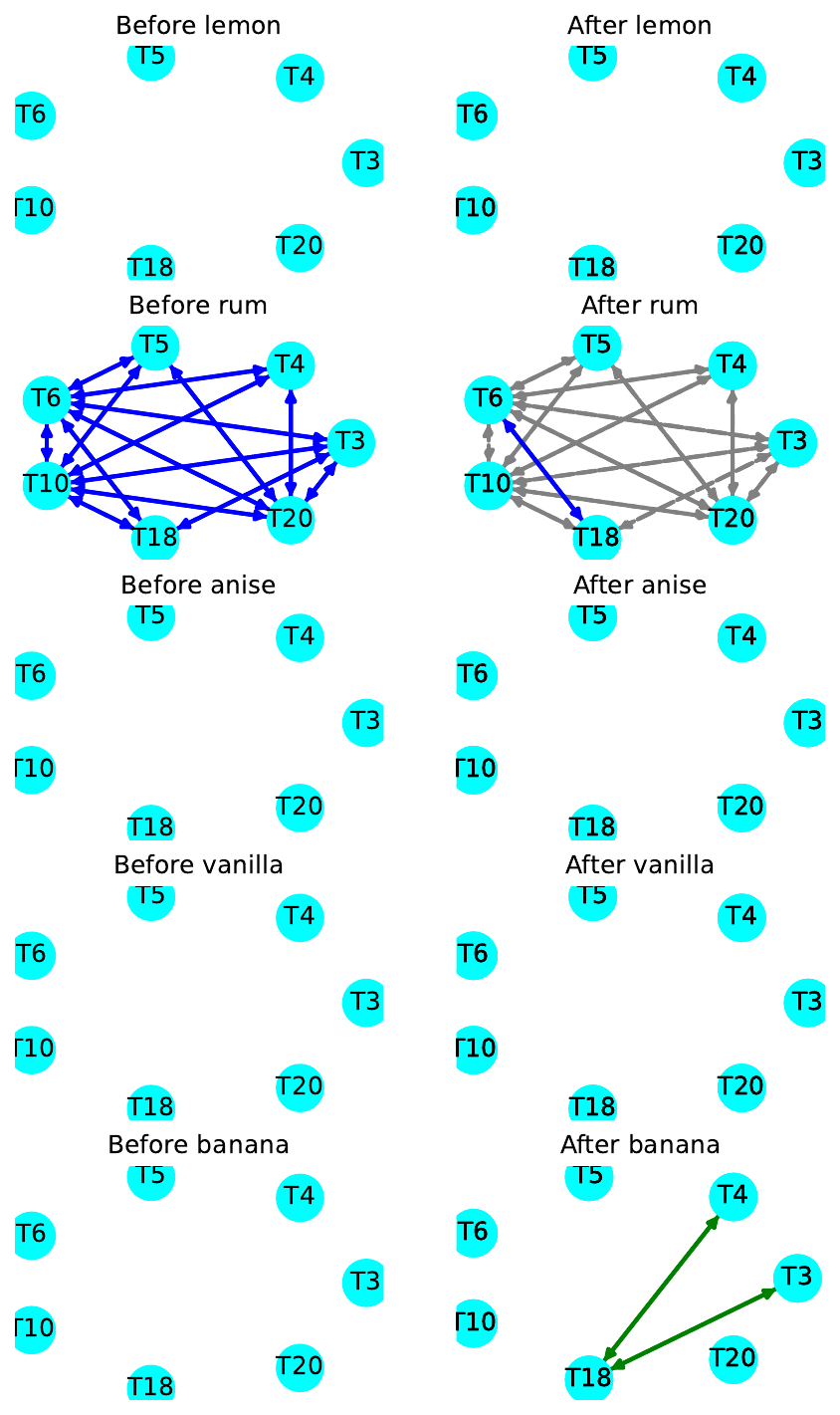}
    \caption{In the prior to odor presentation networks (based on the realizations at the start of the epoch) an edge is present if the high frequency coherence connectivity is higher than some threshold as defined by the 0.9 quantile. On the post odor presentation networks (based on the realizations at the start of the epoch), a blue edge indicates that this connectivity was high prior to and it remained post odor presentation, a dotted line gray edge indicates that this connectivity was high before and it is no longer the case and finally a green edge indicates that this connectivity was low before and it became high after the introduction of the smell.}
    \label{fig:network_coherence_low_freq}
\end{figure}
\begin{figure}[H]
    \centering
    \includegraphics[width=.99\linewidth]{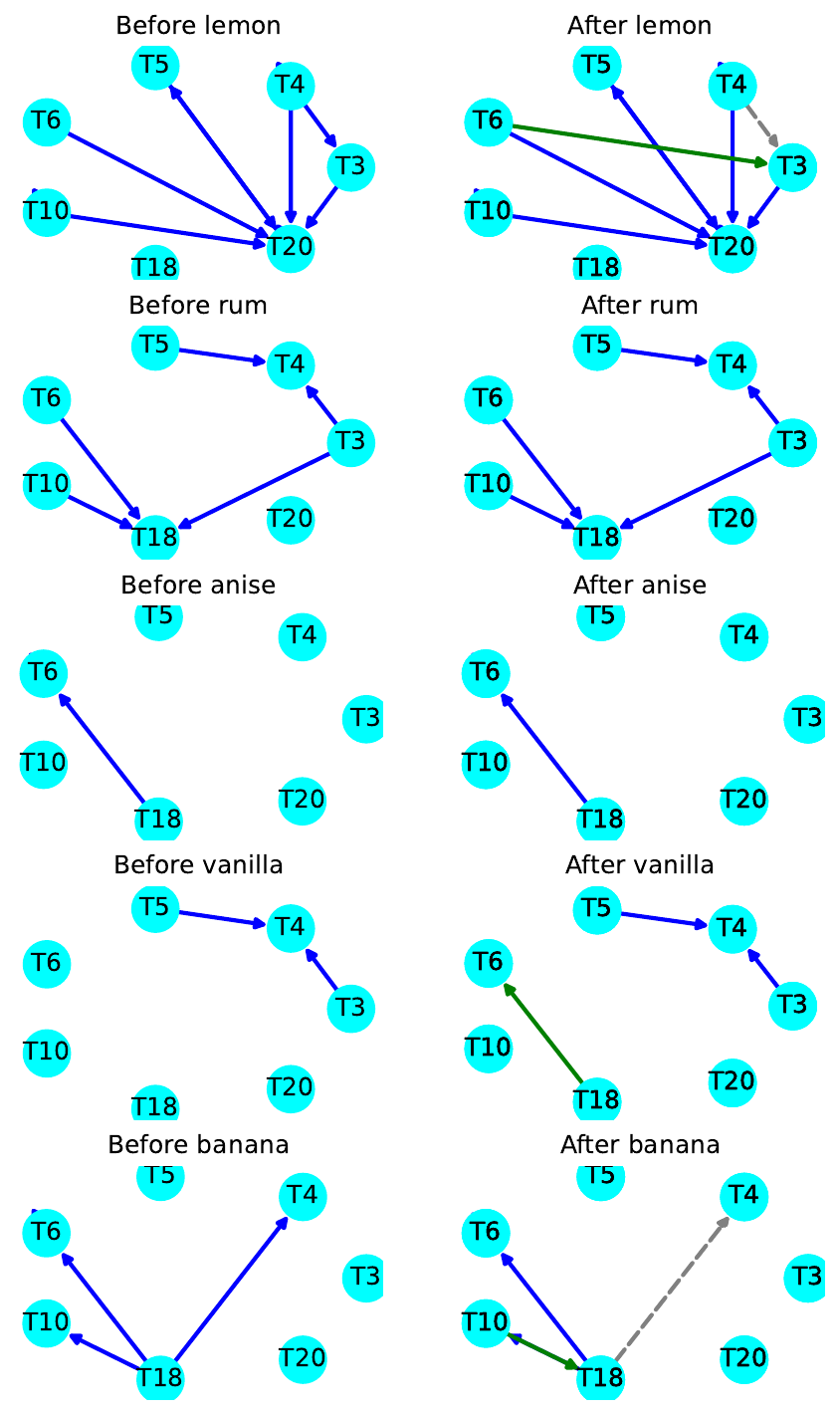}
    \caption{In the prior to odor presentation networks (based on the realizations at the start of the epoch) an edge is present if the low frequency PDC connectivity is higher than some threshold as defined by the 0.9 quantile. On the post odor presentation networks (based on the realizations at the start of the epoch), a blue edge indicates that this connectivity was high prior to and it remained post odor presentation, a dotted line gray edge indicates that this connectivity was high before and it is no longer the case and finally a green edge indicates that this connectivity was low before and it became high after the introduction of the smell.}
    \label{fig:network_pdc_low_freq}
\end{figure}


%

\newpage

\bibliographystyle{imsart-nameyear} 

\bibliography{references}   
\end{document}